\input harvmac
\input epsf
\input amssym
%
\noblackbox

\newcount\figno
\figno=0
\def\fig#1#2#3{
\par\begingroup\parindent=0pt\leftskip=1cm\rightskip=1cm\parindent=0pt
\baselineskip=11pt
\global\advance\figno by 1
\midinsert
\epsfxsize=#3
\centerline{\epsfbox{#2}}
\vskip -21pt
{\bf Fig.\ \the\figno: } #1\par
\endinsert\endgroup\par
}
\def\figlabel#1{\xdef#1{\the\figno}}
\def\encadremath#1{\vbox{\hrule\hbox{\vrule\kern8pt\vbox{\kern8pt
\hbox{$\displaystyle #1$}\kern8pt}
\kern8pt\vrule}\hrule}}

\ifx\pdfoutput\undefined
\def\rgboo#1{}
\else
\def\rgboo#1{\pdfliteral{#1 rg #1 RG}}

\fi
%
\def\Black{\rgboo{0 0 0}}
\def\White{\rgboo{1 1 1}}



\def\frac#1#2{{#1 \over #2}}

\def\semi{\subset\kern-1em\times\;}

\def\sqr#1#2{{\vcenter{\vbox{\hrule height.#2pt
\hbox{\vrule width.#2pt height#1pt \kern#1pt \vrule width.#2pt}
\hrule height.#2pt}}}}

     \def\cO{{\cal O}}
%

%

%

%

%

%

%
\def\coeff#1#2{\relax{\textstyle {#1 \over #2}}\displaystyle}
\def\ZZ{\Bbb{Z}}

\def\IR{\Bbb{R}}

\lref\GiustoID{
  S.~Giusto, S.~D.~Mathur and A.~Saxena,
  ``Dual geometries for a set of 3-charge microstates,''
  Nucl.\ Phys.\ B {\bf 701}, 357 (2004)
  [arXiv:hep-th/0405017].
}

\lref\BenaWV{
  I.~Bena,
  ``Splitting hairs of the three charge black hole,''
  Phys.\ Rev.\ D {\bf 70}, 105018 (2004)
  [arXiv:hep-th/0404073].
}

\lref\BenaDE{
I.~Bena and N.~P.~Warner,
``One ring to rule them all ... and in the darkness bind them?,''
Adv. Theor. Math. Phys. 9 (2006) 1-35
[arXiv:hep-th/0408106].
}
%
\lref\BenaTD{
  I.~Bena, C.~W.~Wang and N.~P.~Warner,
``Black rings with varying charge density,''
  HEP {\bf 0603}, 015 (2006)
  [arXiv:hep-th/0411072].
}
\lref\BenaVA{
  I.~Bena and N.~P.~Warner,
``Bubbling supertubes and foaming black holes,''
  arXiv:hep-th/0505166.
}
%
\lref\BerglundVB{
  P.~Berglund, E.~G.~Gimon and T.~S.~Levi,
``Supergravity microstates for BPS black holes and black rings,''
  arXiv:hep-th/0505167.
}
%
\lref\BenaZY{
  I.~Bena, C.~W.~Wang and N.~P.~Warner,
``Sliding rings and spinning holes,''
  arXiv:hep-th/0512157.
 }
%
\lref\GiustoIP{
  S.~Giusto, S.~D.~Mathur and A.~Saxena,
 ``3-charge geometries and their CFT duals,''
  Nucl.\ Phys.\ B {\bf 710}, 425 (2005)
  [arXiv:hep-th/0406103].
}
%
\lref\GiustoKJ{
  S.~Giusto and S.~D.~Mathur,
   ``Geometry of D1-D5-P bound states,''
  Nucl.\ Phys.\ B {\bf 729}, 203 (2005)
  [arXiv:hep-th/0409067].
}
\lref\StromingerSH{
  A.~Strominger and C.~Vafa,
  ``Microscopic Origin of the Bekenstein-Hawking Entropy,''
  Phys.\ Lett.\ B {\bf 379}, 99 (1996)
  [arXiv:hep-th/9601029].
}

\lref\BreckenridgeIS{
J.~C.~Breckenridge, R.~C.~Myers, A.~W.~Peet and C.~Vafa,
``D-branes and spinning black holes,''
Phys.\ Lett.\ B {\bf 391}, 93 (1997)
[arXiv:hep-th/9602065].
}
%
\lref\LuninJY{
  O.~Lunin and S.~D.~Mathur,
  ``AdS/CFT duality and the black hole information paradox,''
  Nucl.\ Phys.\ B {\bf 623}, 342 (2002)
  [arXiv:hep-th/0109154].
}
\lref\LuninIZ{
  O.~Lunin, J.~M.~Maldacena and L.~Maoz,
  ``Gravity solutions for the D1-D5 system with angular momentum,''
  arXiv:hep-th/0212210.
}

\lref\GauntlettNW{
J.~P.~Gauntlett, J.~B.~Gutowski, C.~M.~Hull, S.~Pakis and
H.~S.~Reall,
``All supersymmetric solutions of minimal supergravity
 in five dimensions,''
Class.\ Quant.\ Grav.\  {\bf 20}, 4587 (2003)
[arXiv:hep-th/0209114].
%
}
\lref\GutowskiYV{
J.~B.~Gutowski and H.~S.~Reall,
``General supersymmetric AdS(5) black holes,''
JHEP {\bf 0404}, 048 (2004)
[arXiv:hep-th/0401129].
}
%
\lref\BenaTK{
  I.~Bena and P.~Kraus,
``Microscopic description of black rings in AdS/CFT,''
  JHEP {\bf 0412}, 070 (2004)
  [arXiv:hep-th/0408186].
}
\lref\BenaWT{
I.~Bena and P.~Kraus,
``Three charge supertubes and black hole hair,''
Phys.\ Rev.\ D {\bf 70}, 046003 (2004)
[arXiv:hep-th/0402144].
}
\lref\GauntlettQY{
  J.~P.~Gauntlett and J.~B.~Gutowski,
``General concentric black rings,''
  Phys.\ Rev.\ D {\bf 71}, 045002 (2005)
  [arXiv:hep-th/0408122].
}

\lref\ElvangDS{H.~Elvang, R.~Emparan, D.~Mateos and H.~S.~Reall,
``Supersymmetric black rings and three-charge supertubes,''
  Phys.\ Rev.\ D {\bf 71}, 024033 (2005)
  [arXiv:hep-th/0408120].
}
\lref\ElvangRT{H.~Elvang, R.~Emparan, D.~Mateos and H.~S.~Reall,
``A supersymmetric black ring,''
  Phys.\ Rev.\ Lett.\  {\bf 93}, 211302 (2004)
  [arXiv:hep-th/0407065].
}
%
\lref\PalmerGU{
  B.~C.~Palmer and D.~Marolf,
``Counting supertubes,''
  JHEP {\bf 0406}, 028 (2004)
  [arXiv:hep-th/0403025].
}
\lref\GreenSP{
  M.~B.~Green, J.~H.~Schwarz and E.~Witten,
``Superstring Theory. Vol. 1: Introduction,''
}
%
\lref\DenefRU{
  F.~Denef,
``Quantum quivers and Hall/hole halos,''
  JHEP {\bf 0210}, 023 (2002)
  [arXiv:hep-th/0206072].
}
%
\lref\DenefNB{
  F.~Denef,
``Supergravity flows and D-brane stability,''
  JHEP {\bf 0008}, 050 (2000)
  [arXiv:hep-th/0005049].
}
%
\lref\BatesVX{
  B.~Bates and F.~Denef,
``Exact solutions for supersymmetric stationary black hole composites,''
  arXiv:hep-th/0304094.
}
%
\lref\KalloshVY{
  R.~Kallosh, A.~Rajaraman and W.~K.~Wong,
``Supersymmetric rotating black holes and attractors,''
  Phys.\ Rev.\ D {\bf 55}, 3246 (1997)
  [arXiv:hep-th/9611094].
}
\lref\MathurZP{
  S.~D.~Mathur,
``The fuzzball proposal for black holes: An elementary review,''
  Fortsch.\ Phys.\  {\bf 53}, 793 (2005)
  [arXiv:hep-th/0502050].
}
%
\lref\GrantQC{
  L.~Grant, L.~Maoz, J.~Marsano, K.~Papadodimas and V.~S.~Rychkov,
``Minisuperspace quantization of 'bubbling AdS' and free fermion droplets,''
  JHEP {\bf 0508}, 025 (2005)
  [arXiv:hep-th/0505079].
}
%
\lref\BenaAY{
  I.~Bena and P.~Kraus,
``Microstates of the D1-D5-KK system,''
  Phys.\ Rev.\ D {\bf 72}, 025007 (2005)
  [arXiv:hep-th/0503053].
}
\lref\GiustoZI{
  S.~Giusto, S.~D.~Mathur and Y.~K.~Srivastava,
``A microstate for the 3-charge black ring,''
  arXiv:hep-th/0601193.
}
%
\lref\SaxenaUK{
  A.~Saxena, G.~Potvin, S.~Giusto and A.~W.~Peet,
``Smooth geometries with four charges in four dimensions,''
  arXiv:hep-th/0509214.
}
%
\lref\RychkovJI{
  V.~S.~Rychkov,
``D1-D5 black hole microstate counting from supergravity,''
  JHEP {\bf 0601}, 063 (2006)
  [arXiv:hep-th/0512053].
}
%
%
\lref\GibbonsSP{
  G.~W.~Gibbons and P.~J.~Ruback,
``The Hidden Symmetries Of Multicenter Metrics,''
  Commun.\ Math.\ Phys.\  {\bf 115}, 267 (1988).
}
%
\lref\HitchinEA{
  N.~J.~Hitchin, A.~Karlhede, U.~Lindstrom and M.~Rocek,
``Hyperkahler Metrics And Supersymmetry,''
  Commun.\ Math.\ Phys.\  {\bf 108}, 535 (1987).
}
\lref\KarlhedeMG{
  A.~Karlhede, U.~Lindstrom and M.~Rocek,
``Hyperkahler Manifolds And Nonlinear Supermultiplets,''
  Commun.\ Math.\ Phys.\  {\bf 108}, 529 (1987).
}
%
\lref\LindstromKS{
  U.~Lindstrom and M.~Rocek,
``New Hyperkahler Metrics And New Supermultiplets,''
  Commun.\ Math.\ Phys.\  {\bf 115}, 21 (1988).
}
%
\lref\IvanovCY{
  I.~T.~Ivanov and M.~Rocek,
``Supersymmetric sigma models, twistors, and the Atiyah-Hitchin metric,''
  Commun.\ Math.\ Phys.\  {\bf 182}, 291 (1996)
  [arXiv:hep-th/9512075].
}
%
\lref\BenaIS{
 I.~Bena, C.~W.~Wang and N.~P.~Warner,
``The foaming three-charge black hole,''
arXiv:hep-th/0604110.
}
\lref\frascati{ I.~Bena, Lectures given at  ``Winter School on Attractor Mechanism,'' INFN-Laboratori Nazionali di Frascati, Italy, March 2006, to appear.}

\lref\BoyerMM{
  C.~P.~Boyer and J.~D.~.~Finley,
  ``Killing Vectors In Selfdual, Euclidean Einstein Spaces,''
  J.\ Math.\ Phys.\  {\bf 23}, 1126 (1982).
}

\lref\gegenberg{
J.~Gegenberg and A.~Das,
``Stationary Riemannian space-times with self-dual curvature,''
 Gen.\ Rel.\ Grav. {\bf 16} (1984) 817.
}
%

\lref\AharonyTI{
  O.~Aharony, S.~S.~Gubser, J.~M.~Maldacena, H.~Ooguri and Y.~Oz,
  ``Large N field theories, string theory and gravity,''
  Phys.\ Rept.\  {\bf 323}, 183 (2000)
  [arXiv:hep-th/9905111].
}
\lref\DavidWN{
  J.~R.~David, G.~Mandal and S.~R.~Wadia,
  ``Microscopic formulation of black holes in string theory,''
  Phys.\ Rept.\  {\bf 369}, 549 (2002)
  [arXiv:hep-th/0203048].
}

\lref\MaldacenaDS{
  J.~M.~Maldacena and L.~Susskind,
  ``D-branes and Fat Black Holes,''
  Nucl.\ Phys.\ B {\bf 475}, 679 (1996)
  [arXiv:hep-th/9604042].
}

\lref\BalasubramanianGI{
  V.~Balasubramanian, E.~G.~Gimon and T.~S.~Levi,
  ``Four dimensional black hole microstates: From D-branes to spacetime foam,''
  arXiv:hep-th/0606118.
}
\lref\Saxena{A.~Saxena, to appear.}

\lref\GauntlettWH{
  J.~P.~Gauntlett and J.~B.~Gutowski,
  ``Concentric black rings,''
  Phys.\ Rev.\ D {\bf 71}, 025013 (2005)
  [arXiv:hep-th/0408010].
}

\lref\LuninUU{
  O.~Lunin,
  ``Adding momentum to D1-D5 system,''
  JHEP {\bf 0404}, 054 (2004)
  [arXiv:hep-th/0404006].
}



\Title{
\vbox{
\hbox{\baselineskip12pt \vbox{\hbox{hep-th/0608217}
\hbox{NSF-KITP-06-26}
\hbox{CERN-PH-TH/2006-151}}
}}}
{\vbox{\vskip -1.5cm
\centerline{\hbox{Mergers and Typical  Black Hole Microstates }}
}}
\vskip -.3cm
\centerline{Iosif~Bena${}^{(1)}$,  Chih-Wei Wang${}^{(2)}$ and
Nicholas P.\ Warner${}^{(2,3)}$}

\bigskip
\centerline{{${}^{(1)}$\it School of Natural Sciences,
Institute for Advanced Study }}
\centerline{{\it Einstein Dr., Princeton, NJ 08540, USA }}
\medskip
\centerline{{${}^{(2)}$\it Department of Physics and Astronomy,
University of Southern California}} \centerline{{\it Los Angeles,
CA 90089-0484, USA}}
\medskip
\centerline{{${}^{(3)}$\it Department of Physics, Theory Division}}
\centerline{{\it CERN, Geneva, Switzerland}}
\medskip

\bigskip
\bigskip

We use mergers of microstates to obtain the first smooth horizonless
microstate solutions corresponding to a BPS three-charge black hole with a
classically large horizon area. These microstates have very long
throats, that become infinite in the classical limit; nevertheless,
their curvature is everywhere small. Having a classically-infinite
throat makes these microstates very similar to the typical microstates
of this black hole. A rough CFT analysis confirms this intuition, and
indicates a possible class of dual CFT microstates.

We also analyze the properties and the merging of microstates
corresponding to zero-entropy BPS black holes and black rings. We find
that these solutions have the same size as the horizon size of their
classical counterparts, and we examine the changes of internal structure
of these microstates during mergers.

\vskip .3in
\Date{\sl {August 2006}}

\vfill\eject

\newsec{Introduction}

In  \BenaIS\ we found a very large number of geometries that have
exactly the same supersymmetries, size, charges and angular momenta as
maximally-rotating, three-charge BPS black holes and black rings in
five dimensions. These smooth, horizonless solutions have no brane
sources, and belong to the class constructed and analyzed in \refs{
\BenaVA, \BerglundVB}, based on earlier work in
\refs{\GauntlettNW\GutowskiYV\BenaDE-\GauntlettQY}.

These solutions are very similar to BPS black holes and BPS black rings
\refs{\BenaWT,\ElvangRT,\BenaDE,\ElvangDS,\GauntlettQY} and
only differ from those near the would-be ``classical'' horizon. In the
new, horizonless solutions this region is smooth and compact,
and contains a large number of topologically non-trivial
two-cycles.  Hence these solutions are called ``bubbled black holes''
and ``bubbled black rings.''  All these solutions can be dualized to a
frame in which they are asymptotic to $AdS_3 \times S^3 \times T^4$,
and therefore are dual to states in the D1-D5-P CFT that describes
black holes and black rings.  Understanding whether this kind of
supergravity solutions are dual to {\it typical} CFT states
is perhaps among the most important questions in understanding black holes
in string theory\foot{If the answer to this question is positive, then the $AdS$-CFT
correspondence would force us to think about black holes as ensembles
of smooth horizonless geometries, which would greatly deepen our
understanding of black holes, and quantum gravity in general. A review
of this can be found in \MathurZP\ and \frascati.}.

For practical reasons, all the five-dimensional, three-charge BPS
microstates that have been constructed \refs{\LuninUU\GiustoID\GiustoIP\GiustoKJ-\GiustoZI,\BenaIS,\BenaVA,\BerglundVB} use as a base a
four-dimensional ``generalized hyper-K\"ahler'' metric\foot{By
``generalized,'' we mean a hyper-K\"ahler base whose metric is allowed to
change its overall sign in compact regions, thereby flipping the
signature from $+4$ to $-4$.} that has a tri-holomorphic $U(1)$
symmetry, and is thus a Gibbons-Hawking (GH) metric.  Nevertheless, as recently
shown in \BenaIS, using a Gibbons-Hawking base generically appears to yield
bubbled solutions corresponding to black rings and black holes of zero
horizon area.  Indeed, it appears that {\it none} of the microstate
solutions currently in the literature has the same charges and angular momenta
as a black hole with a classically large horizon area.

It is very important to understand whether this limitation arises from
using a Gibbons-Hawking base.  If this were so, then we would need to
find more general classes of hyper-K\"ahler base metrics, and these
are very hard to construct and analyze explicitly. Before embarking
on this difficult task, one should therefore first attempt to
construct microstates of positive-entropy black holes starting from a
Gibbons-Hawking base.    Again this is primarily because of
computational convenience:  Such solutions would be much easier to obtain
explicitly, analyze, and relate to the corresponding CFT microstates.

An obvious place to start would be to examine the merger of two  black holes,
which is always irreversible.  Nevertheless, such a configuration can preserve at most and $SO(3)$ that  has no tri-holomorphic $U(1)$ subgroups; therefore the merger of two black hole microstates  cannot be described using a
GH base metric.
On the other hand, the merger of a black hole with
a black ring in its equatorial plane preserves a $U(1) \times U(1)$
symmetry.  Hence, using a result of \GibbonsSP, we expect the merger of the corresponding microstates  to be described
using a GH base metric.

The merger of a BPS black hole and a BPS  black ring  can be reversible or
irreversible, depending on the charges of the two objects \BenaZY. One therefore
expects the merger of microstates to result in an zero-entropy BH microstate or an
positive-entropy BH microstate\foot{Obviously, the compound adjectives,
``positive-entropy'' and ``zero-entropy''
are intended to be applied to the black hole and black ring whose microstate
geometries we discuss, and {\it not} to the horizonless microstate geometries
themselves.}, depending on the charges of the merging
microstates. Moreover, since the merger can be achieved by keeping the
base Gibbons-Hawking, one expects the resulting positive-entropy BH microstate to
have a Gibbons-Hawking base.

The main purpose of this paper is to study the merger of zero-entropy BR
microstates and zero-entropy BH microstates. We construct solutions that
describe a large, bubbled black ring with a bubbling black hole in the
center, and reduce some flux parameters in order to bring them
together.  We find that, when seen from far away, the merger of
microstates closely parallels the merger of the corresponding classical
counterparts.  In particular, the merger happens at the same values of
the charges and angular momenta, and the resulting microstate is
always that of a BMPV black hole.

On the other hand, we find that there is a huge qualitative difference
between the behavior of the internal structure of microstates in
``reversible'' and ``irreversible'' mergers\foot{With an obvious abuse
of terminology, we will refer to such solutions as ``reversible'' and
``irreversible'' mergers of microstates with the understanding that
the notion of reversibility refers to the classical BH and BR
solutions to which the microstates correspond.}. A ``reversible''
merger of an zero-entropy BH microstate and  an zero-entropy BR microstate
produces another zero-entropy BH microstate. We find that the bubbles
corresponding to the ring simply join the bubbles corresponding to the
black hole, and form a bigger bubbled structure.

In an ``irreversible'' merger, as the ring bubbles and the black
hole bubbles get closer and closer, we find that the distances between the GH
points that form the black hole foam and the black ring foam also
decrease. As one approaches the merger point, all the sizes in the
GH base scale down to zero while preserving their relative proportions.
In the limit in which the merger occurs, the solutions have $J_1=J_2 <
\sqrt{Q_1 Q_2 Q_3}$, and all the points have scaled down to zero
size on the base.   Therefore, it naively looks like the
configuration is singular; however, the full physical size of the
bubbles also depends on the warp factors, and taking these into
account one can show that the physical size of all the bubbles in the black hole and black ring foams remains the same. The
fact that the GH points get closer and closer together implies
that the throat of the solution becomes deeper and deeper, and
more and more similar to the throat of a BPS black hole (which is
infinite).

\goodbreak\midinsert
\vskip -.02cm
\centerline{ {\epsfxsize 4.8in\epsfbox{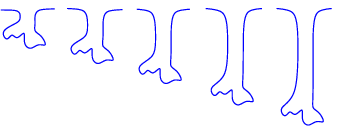}}}
\vskip 0.2cm
\leftskip 2pc
\rightskip 2pc\noindent{\ninepoint\sl \baselineskip=8pt
{\bf Fig.~1}:
A heuristic depiction of an irreversible microstate merger. Throughout the merger the
physical size of the bubbles remains fixed, but the throat of the solutions deepens. }

\endinsert

Our investigation therefore shows that Gibbons-Hawking-base microstates
of positive-entropy black holes can be obtained from mergers. Moreover, these
microstates appear to want to have very long throats, and consequently
very high red-shifts. As we will see, the quantization of the fluxes
does not allow the scaling to continue {\it ad infinitum}; the
length of the throats of microstate solutions only becomes infinite in the
classical limit.

Hence, at least for a Gibbons-Hawking base, the
crucial difference between zero-entropy BH microstates and positive-entropy BH
microstates obtained from mergers is the length of the throat. The throat of the
zero-entropy BH microstates remains finite in the classical limit, while the throat of the
positive-entropy BH microstates becomes infinite. In the
rest of this paper we will refer to the microstates of positive-entropy BH's as
``deep'' microstates\foot{We originally considered calling these
microstates ``deep-throat microstates'' but decided against it to
avoid the obvious tasteless jokes.}, and to the  zero-entropy BH microstates
as ``shallow'' microstates.

As we have mentioned above, all the solutions we construct are dual
to states of the D1-D5-P CFT. A heuristic way to describe the
states of this system is to consider a long effective
string of total length $N_1 N_5$ that is broken into component strings,
 on which one puts $N_P$ units of
momentum. The CFT entropy comes from the various ways of distributing
the momentum between the different modes of the component strings. The typical
CFT states, dual to the black hole, have only one component string, that carries
all the  momentum; atypical states have more component strings
\refs{\StromingerSH,\BreckenridgeIS}.  The energy gap in the CFT is the
inverse of the length of the largest component string. One can also
compute the mass gap in the bubbling solutions, as a function of the
charges and angular momentum, and compare it to the gap computed in
the CFT.  A heuristic analysis then indicates that the deep
microstates with left-moving angular momentum $J_L$ are dual to states
with $J_L$ component strings of length $N_1 N_5 /J_L$. The deepest of
the deep microstates have $J_L=1$, and therefore appear to be dual to
CFT states with a single component string.  While this is not enough to establish that deep microstates are typical -- after all, they
might be dual to an atypical distribution of momentum on the component string -- the fact that they are dual to CFT states with one component string makes them very similar to typical microstates.

In addition to obtaining microstates corresponding to black holes with
classically large entropy, we can use mergers to investigate the
``size'' of microstates.  When a classical black hole and a classical
black ring merge, their horizons touch.  Our results indicate that in
a similar fashion, when a bubbled black hole and a bubbled black ring
merge, their bubbles get very close to each other. This shows that the bubbling black
hole microstates constructed in \BenaIS\ have the same size and macroscopic
features as the corresponding classical black hole.

Another purpose of this paper is to discuss, in more detail, the
construction of bubbled  black holes and black rings outlined in
\BenaIS, and to analyze more thoroughly the entropy of the bubbled
black holes and black rings with a GH space. Our analysis of
mergers can also be used to find the detailed features of the
distribution of the GH points that form bubbled black holes and
bubbled black rings, as well as the behavior of the warp factors
inside these solutions, and is important for obtaining a deeper
understanding of their physics.

In section 2 we summarize the structure and properties of the five
dimensional smooth horizonless solutions of \refs{\BenaVA,
\BerglundVB}. In section 3 we give an expanded version of the
construction of bubbling black holes and black rings in \BenaIS.  We
 also give some numerical results that suggest how the GH centers
 arrange themselves within the bubbling solutions.  In section 4 we
 analyze the merger of microstates analytically, and show that one can
 rederive the merger conditions found in \BenaZY\ for classical rings and holes,  using the bubbled geometries.  In section 5 we analyze
 analytically and numerically the irreversible mergers, which result
 in positive-entropy BH microstates. These, to our knowledge, are the first
 smooth horizonless solutions that have the same charges and angular momenta as a
 black hole with a classically large horizon area.  In section 6
 we look at some examples of reversible mergers and
 show that during such mergers the separation of the ring and the
 black hole bubbles becomes very small, resulting in a configuration
 where the two bubbling structures essentially touch.  In section 7 we
 make a heuristic $AdS$-CFT analysis of the deep microstates, and
 argue that the deepest microstates correspond to CFT states that have a single component string; these microstates are therefore in the
 same class as the typical microstates of this black hole.
 Section 8 contains conclusions and suggestions for future research.

\newsec{A review of the bubbling solutions}

\subsec{The solution}

The full eleven-dimensional metric of a BPS black-hole and
black-ring solutions has the form:
\eqn\fullmet{\eqalign{ ds_{11}^2& =  - \left({1 \over Z_1 Z_2
Z_3}\right)^{2/3} (dt+k)^2 + \left( Z_1 Z_2 Z_3\right)^{1/3}
h_{mn}dx^m dx^n \cr &+ \left({Z_2 Z_3 \over
Z_1^2}\right)^{1/3}(dx_1^2+dx_2^2) + \left({Z_1 Z_3 \over
Z_2^2}\right)^{1/3}(dx_3^2+dx_4^2) + \left({Z_1 Z_2 \over
Z_3^2}\right)^{1/3}(dx_5^2+dx_6^2) \,,}}
and the standard approach to their constructing
starts by taking the four-dimensional spatial base metric $h_{mn}$
to be that of flat $\IR^4$  \refs{\BenaWV,\ElvangRT,\GauntlettWH,\BenaDE,\ElvangDS,\GauntlettQY }.

The corresponding ``bubbling'' solutions  \refs{\BenaVA,\BerglundVB}
are obtained by replacing this flat $\IR^4$ by a half-flat,  four-dimensional
``generalized hyper-K\"ahler'' metric that is asymptotic to
$\IR^4$.   By ``generalized,'' we mean that one can
use a hyper-K\"ahler base whose metric is allowed to change its overall sign
in compact regions, thereby flipping the signature from $+4$ to $-4$.
The overall, eleven-dimensional  metric is perfectly regular and
Lorentzian essentially because any overall change of sign in the
four-dimensional base metric is exactly compensated by sign
changes in the warp factors $Z_i$. \refs{\GiustoKJ,\BenaVA,\BerglundVB}.

This opens up a huge array of new possibilities for constructing
bubbling black holes and rings, because there are many, many
hyper-K\"ahler metrics that are asymptotic to $\IR^4$ but have
compact regions where the metric changes its overall
sign\foot{Based on the physics of three-charge supertubes \BenaWT,
one expects generalized hyper-K\"ahler metrics with $N$ regions of
negative signature to be parameterized by $6N$ functions of one
variable \BenaVA.}.   For practical reasons, the solutions of
\refs{\BenaIS,\BenaVA,\BerglundVB} were restricted to the simplest class
of generalized hyper-K\"ahler geometries:
Gibbons-Hawking (GH) metrics whose potential, $V$, is allowed to be
negative.   Mathematically, these metrics are extremely special:  They constitute
the complete family of four-dimensional hyper-K\"ahler metrics with a
tri-holomorphic $U(1)$ isometry.  The general  BPS solutions with a
{\it positive-definite} GH base were obtained in \refs{\GauntlettNW,\GauntlettQY},
and it is straightforward
to adapt these solutions to generalized GH metrics that allow overall sign changes.
In this paper we will work entirely within this class of solutions.

We therefore consider a four-dimensional base metric that
has Gibbons-Hawking (GH) form:
\eqn\GHmetric{ds_4^2 ~=~  V^{-1} \, \big( d\psi + \vec{A} \cdot
d\vec{y}\big)^2  ~+~ V\, (d\vec{y}\cdot d\vec{y} )\,,}
where $\vec y \in \IR^3$ and
\eqn\Vform{V ~=~   \sum_{j=1}^N \,  {q_j  \over r_j} \,, \qquad
\vec \nabla \times \vec A ~=~ \vec \nabla V \,,}
with  $r_j \equiv |\vec{y}-\vec{y}^{(j)}|$.   In order for the GH metric
to be regular, one must take $q_j \in \ZZ$ and for the metric to be asymptotic
to that of flat $\IR^4$ one must also impose
\eqn\qzero{q_0 ~\equiv~ \sum_{j=1}^N \, q_j  ~=~ 1\,.}
The GH metric has non-trivial two-cycles (the {\it bubbles}), $\Delta_{ij}$,
defined by the fiber  coordinate, $\psi$, and any line running between a pair of
GH points, $\vec{y}^{(i)}$ and  $\vec{y}^{(j)}$.

The background three-form potentials in the eleven-dimensional solution
are defined via vector potentials in the four-dimensional GH base.
\eqn\Aansatz{
 {\cal A}   ~=~  A^{(1)} \wedge dx_1 \wedge dx_2 ~+~  A^{(2)}   \wedge
dx_3
 \wedge dx_4 ~+~ A^{(3)}  \wedge dx_5 \wedge dx_6\,,}
Supersymmetry requires the  ``dipole field strengths:''
\eqn\Thetadefn{\Theta^{(I)} \equiv d A^{(I)} + d\Big(  {dt +k \over
Z_I}\Big)}
to be self-dual. One can easily show that if the base has a
Gibbons-Hawking metric, then the   $\Theta^I$ are given by
\eqn\thetaansatz{\Theta^{(I)} ~=~ - \, \sum_{a=1}^3 \,
\big[\partial_a \big( V^{-1}\, K^I \big)\big] \, \big[ (d \psi
+A)\wedge  dy^a + \coeff{1}{2}\, V \epsilon_{abc} \, dy^b \wedge
dy^c  \big] \,.}
where $K^I$ is harmonic in the $\IR^3$ base of the GH space. To have
 completely non-singular solutions one must choose
these  to be sourced only at the singular points of $V$:
\eqn\KIform{K^I ~=~ \sum_{j=1}^N \, {k_j^I \over r_j} \,.}
One then finds that the two-form fluxes through $\Delta_{ij}$ are given by:
\eqn\fluxperiods{ \Pi^{(I)}_{ij}  ~=~   \bigg( {k_j^I \over q_j} ~-~
{k_i^I \over q_i} \bigg) \,.}

Having chosen the flux parameters, $k_j^I$, the solution is completely
fixed by its asymptotics and by requiring the absence of any singularities
and, in particular, the absence of singular sources.  One introduces functions,
$L^I$, and $M$ defined  by:
\eqn\LMform{ L_I ~=~ 1  ~-~  \half \, C_{IJK}  \, \sum_{j=1}^N \,  {k_j^J \, k_j^K
\over q_j}\, {1 \over r_j} \,, \quad
M ~=~ m_0 ~+~  \coeff{1}{12}\, C_{IJK} \, \sum_{j=1}^N \, {k_j^I \, k_j^J \, k_j^K
\over q_j^2} \,  {1 \over r_j} \,,}
with
\eqn\mzerodefn{m_0  ~=~  -\half \, q_0^{-1} \, \sum_{j=1}^N\, \sum_I \, k_j^I  ~=~
 -\half \,  \sum_{j=1}^N\, \sum_I \, k_j^I   \,.}

The warp factors, $Z_I$, are then given by:
\eqn\ZIform{Z_I ~=~  L_I ~+~ \half \, C_{IJK} \, V^{-1}\,K^J K^K \,,}
and the angular momentum vector, $k$, in \fullmet\ is:
\eqn\kform{k ~=~ \mu\, ( d\psi + A   ) ~+~ \omega \,,}
with
\eqn\muform{\mu ~=~ \coeff {1}{6} \, V^{-2} \,C_{IJK}\,   K^I K^J K^K  ~+~
\coeff{1}{2}\, V^{-1} \, K^I L_I ~+~  M\,,}
and $\omega$ defined by $\omega = \vec \omega \cdot d\vec y$, where
\eqn\omegeqn{\vec \nabla \times \vec \omega ~=~  V \vec \nabla M ~-~
M \vec \nabla V ~+~   \coeff{1}{2}\, (K^I  \vec\nabla L_I - L_I \vec
\nabla K^I )\,.}
One needs to solve this equation without introducing Dirac-Misner
strings into the metric  and to do this it is convenient to introduce
one forms, $\omega_{ij}$, associated to each pair of GH points.
The simplest way to define these forms is to choose coordinates so that
$\vec y = (x,y,z)$ and  $\vec y^{(i)} = (0,0,a)$ and $\vec y^{(j)} = (0,0,b)$,
with $a>b$, and then one sets:
\eqn\omegaijdefn{ \omega_{ij} ~\equiv~
 -  {(x^2 +  y^2 + (z-a+ r_i)(z-b - r_j)) \over (a-b)  \, r_i  \, r_j } \, d \phi  \,,}
where $\tan \phi = y/x$.   The importance of this form is that it has
no Dirac strings.   The desired non-singular solution to \omegeqn\
may then be written as
\eqn\omegacomplete{\eqalign{\vec  \omega    ~=~  \coeff{1}{24} \,
C_{IJK} \,   \sum_{i, j =1}^N \, q_i \, q_j \,   \Pi^{(I)}_{ij} \,  \Pi^{(J)}_{ij} \,
\Pi^{(K)}_{ij} \,  \vec \omega_{ij}  \,,}}
provided that  the {\it bubble equations} are satisfied:
\eqn\bubbleeqns{ \coeff{1}{6}\, C_{IJK} \, \sum_{{\scriptstyle j=1} \atop
{\scriptstyle j \ne i}}^N \,  \,  \Pi^{(I)}_{ij} \,   \Pi^{(J)}_{ij} \,  \Pi^{(K)}_{ij} \
{q_i \, q_j  \over r_{ij} } ~=~ -2\, \Big(m_0 \, q_i ~+~  \half
\sum_{I=1}^3  k^I_i \Big) \,, }
for $i =1, \dots, N$, and where $r_{ij} \equiv |\vec y^{(i)} - \vec y^{(j)}|$.
The bubble equations are required to remove  closed timelike
curves (CTC's) in specific, ``potentially dangerous'' limits\foot{One can add a constant to $V$ in \Vform,
and reduce the resulting smooth five-dimensional solution to a singular
four-dimensional multi-black-hole solution of the type explored in
\refs{\DenefNB,\BatesVX}. The  ``bubble equations'' are then equivalent
to the ``integrability conditions'' of \refs{\DenefNB,\BatesVX}. Other asymptotically four-dimensional configurations
that are smooth in five-dimensions and are microstates of four-dimensional black holes have been explored in
\refs{\BenaAY,\SaxenaUK,\BalasubramanianGI}.}.  In general, the absence
of CTC's requires that one ensure that the following are globally true:
\eqn\noCTCs{ V \, Z_I ~ \ge~ 0 \,, \qquad  Z_1\,Z_2 \, Z_3 \, V~-~ \mu^2 \, V^2 ~\ge~ 0 \,,}
However, in quite a number of examples one finds that the bubble
equations suffice to guarantee the global absence of CTC's.

Finally, one should note that shifting $K^I \to K^I + c^I  V$ for some constants,
$c^I$, has a trivial action on the solution and so the parameters, $k^I_j$,
have a gauge invariance:
\eqn\gauge{ k^I_j \rightarrow k^I_j ~+~ q_j \, c^I \,.}
All the physical quantities must be  invariant under this transformation.

\subsec{The charges and angular momenta of the solution}

To isolate the charges of the solution one should first recall that
if $V={1 \over r}$, then the coordinate transformation that takes
\GHmetric\ to a standard polar form for $\IR^4$ is $r = {1 \over 4} \rho^2$,
where $\rho$ is the standard radial coordinate.  Imposing \qzero\ means
that we have $V \sim {1 \over r}$ at infinity, and so this gives us the proper
radial coordinate asymptotically.

To obtain the electric charges measured at infinity, one simply
needs to extract the coefficient of $\rho^{-2}$ in the expansion of
the $Z_I$.  It is elementary to see that:
\eqn\QIchg{Q_I ~=~ -2 \, C_{IJK} \, \sum_{j=1}^N \, q_j^{-1} \,
\tilde  k^J_j \, \tilde  k^K_j\,,}
where
\eqn\ktilde{\tilde  k^I_j ~\equiv~ k^I_j ~-~    q_j\, N  \,  k_0^I  \,,
\qquad {\rm and} \qquad k_0^I ~\equiv~{1 \over N} \, \sum_{j=1}^N k_j^I\,.}
Note that $\tilde  k^I_j$ is invariant under \gauge.

To read off the angular momenta one looks at the asymptotic
behavior of $k$ in \kform\ and extracts the terms that fall-off
as ${ 1 \over \rho^2} \sim  { 1 \over 4 r}$.
Indeed, it is easiest to get the result from the coefficient of $d \psi$,
that is, from the function $\mu$.  There are two types of such terms,
simple ${1 \over r}$ terms and the dipole terms  arising from the expansion
of $V^{-1} K^I $.  Following \BerglundVB, we introduce the dipoles
\eqn\dipoles{\vec D_j ~\equiv~  \, \sum_I  \, \tilde k_j^I \, \vec y^{(j)} \,,
\qquad \vec D ~\equiv~ \sum_{j=1}^N \, \vec D_j \,.}
and then one can obtain  the components of the angular momentum
from:
\eqn\angmomform{k ~\sim~ {1 \over 4 \,\rho^2} \, \big((J_1+J_2) ~+~
(J_1-J_2) \, \cos \theta   \big) \, d\psi ~+~ \dots \,,}
where $\theta$ is the angle between $\vec D$ and $\vec y$.
Expanding the function $\mu$, we find:
\eqn\Jright{ J_R ~\equiv~ J_1 + J_2 ~=~ \coeff{4}{3}\, \, C_{IJK} \, \sum_{j=1}^N q_j^{-2} \,
\tilde  k^I_j \, \tilde  k^J_j \,  \tilde  k^K_j  \,,}
\eqn\Jleft{ J_L ~\equiv~ J_1 - J_2 ~=~ 8 \,\big| \vec D\big|  \,.}
While we have put modulus signs around $\vec D$ in \Jleft, one
should note that it does have a meaningful orientation, and so we
will sometimes consider $\vec J_L = 8 \vec D$.

These results appear to differ by some factors of two compared to those of
\BerglundVB.  This is because our conventions are those of \BenaVA, which
use a different normalization of the two-form fields.

One can use the bubble equations to obtain another, rather more
intuitive expression for $J_1 -J_2$.  One should first note  that the right-hand side
of the bubble equation, \bubbleeqns, may be written as $-  \sum_I  \tilde k_i^I$.
Multiplying this by $\vec y^{(i)}$ and summing over $i$ yields:
\eqn\vecD{\eqalign{\vec J_L ~\equiv~ 8\, \vec D & ~=~  -  \coeff{4}{3}\,
C_{IJK} \, \sum_{{\scriptstyle i, j=1} \atop
{\scriptstyle j \ne i}}^N \,  \,  \Pi^{(I)}_{ij} \,   \Pi^{(J)}_{ij} \,  \Pi^{(K)}_{ij} \
{q_i \, q_j \,  \vec y^{(i)}  \over r_{ij} }
\cr & ~=~-  \coeff{2}{3}\, C_{IJK} \, \sum_{{\scriptstyle i, j=1} \atop
{\scriptstyle j \ne i}}^N \,  q_i \, q_j \,  \Pi^{(I)}_{ij} \,   \Pi^{(J)}_{ij} \,  \Pi^{(K)}_{ij} \
{(\vec y^{(i)} - \vec y^{(j)}) \over \big|\vec y^{(i)} - \vec y^{(j)}\big| } \,,  }}
where we have used the skew symmetry $\Pi_{ij} = - \Pi_{ji}$ to obtain
the second identity.  This result suggests that one should define an angular
momentum flux vector  associated with the $ij^{\rm th}$ bubble:
\eqn\angmomflux{\vec J_{L\, ij} ~\equiv ~ -  \coeff{4}{3}\,q_i \, q_j \, C_{IJK} \,
\Pi^{(I)}_{ij} \,   \Pi^{(J)}_{ij} \,  \Pi^{(K)}_{ij} \, \hat y_{ij} \,, }
where $\hat y_{ij}$ are {\it unit} vectors,
\eqn\unitvecs{\hat y_{ij} ~\equiv ~  {(\vec y^{(i)} - \vec y^{(j)}) \over
 \big|\vec y^{(i)} - \vec y^{(j)}\big| } \,.}
This means that the flux terms
on the left-hand side of the bubble equation actually have a natural spatial
direction, and once this is incorporated, it yields the contribution of the
bubble to $J_L$.

\subsec{The simplest bubbled supertube}

For later convenience we summarize the properties of the solution
with three GH points of charges $q_1=1$, $q_2=-Q$ and $q_3=+Q$ that corresponds to
a zero-entropy black ring \BenaVA.   It is useful to define new,
physical variables $d^I$ and $f^I$ (the $d^I$ are equal to the dipole charges of the ring):
\eqn\dfring{ d^I ~\equiv~ 2 \,(k_2^I ~+~ k_3^I) \,, \qquad
 f_I  ~\equiv~ 2\,  k_1^I ~+~ \big(1+  \coeff{1}{Q} \big)\, k_2^I ~+~
\big (1-  \coeff{1}{Q} \big)\,k_3^I    \,.}
Note that $d^I$ and $f^I$ are invariant under \gauge.

The electric charges of the bubbled tube are:
\eqn\simpringchg{Q_I ~=~ C_{IJK} \, d^J \, f^K\,,}
and the angular momenta are:
\eqn\jonering{   J_1 ~=~  - {  (Q   - 1) \over 12\, Q} \, C_{IJK}\,
{d^I}\,{d^J}\,{d^K}~+~ \half \, C_{IJK}\, {d^I}\,{d^J}\,{f^K} \,,}
\eqn\jtworing{ J_2 ~=~  {{ ( Q -1 ) }^2 \over 24 \, Q^2} \, C_{IJK}\,
{d^I}\,{d^J}\,{d^K} ~+~ \half \,C_{IJK}\, {f^I}\,{f^J} \,{d^K} \,.}
In particular, the angular momentum of the tube is:
\eqn\Jsimptube{ J_T ~=~J_2-J_1~=~  \half\,  C_{IJK}\, ( {f^I}\,{f^J}\,{d^K} -
{d^I}\,{d^J}\,{f^K} ) ~+~   \Big( {3 \,Q^2 - 4\,Q +1 \over 24\, Q^2} \Big)
\, C_{IJK}\, {d^I}\,{d^J}\,{d^K} \,.}
The radius, $R_T$, of the corresponding classical black ring, as measured
in the $\IR^3$ metric of the GH base, can be obtained from:
\eqn\JTRreln{J_T ~=~  4 \, R_T\, (d^1 + d^2 + d^3) \,.}
%

\newsec{Bubbling black holes and rings with a large number of centers}

In \BenaIS\ we constructed bubbled solutions corresponding to maximally-spinning
(zero-entropy) BMPV black holes, or to maximally spinning BPS black rings.
These solutions have a very large number of GH centers and {\it a priori}
there to be rather little difference between bubbling a black hole
and bubbling a black ring: the ring microstates have a blob of GH
centers of zero total charge with a GH  center away from the blob while
the black hole microstates have all the centers in  one blob of net GH charge one.
We will see that this apparently small difference can very significantly influence the
solution of the bubble equations  The purpose of this section is to explain in more
detail the construction of microstates in \BenaIS, and to investigate
numerically the structure of the distribution of GH centers inside the blobs.

\subsec{Microstates of maximally spinning BMPV black holes}

We first consider a configuration of $N$  GH centers that lie is a single ``blob''
and take all these centers to have roughly the same flux parameters, to leading
order in $N$.
To argue that such a blob corresponds to a BMPV black hole, we first need
to show that $J_1 =J_2$.  If the overall configuration has three independent $\ZZ_2$ reflection symmetries then this is trivial because the $\vec D_j$ will then
come in equal and opposite pairs, and so one has $J_L =0$.
More generally, for a ``random'' distribution\foot{Such a distribution
must, of course, satisfy the bubble equations, \bubbleeqns, but this will
still allow a sufficiently random distribution of GH points.}
the vectors $\hat y_{ij}$ will point in ``random'' directions and so
the $\vec J_{L \, ij}$ will generically cancel one another at leading order in $N$.
The only way to  generate a non-zero value of $J_L$ is to bias
the distribution so that there are more positive centers in one region and more
negative ones in another. This is essentially what happens in the microstate
solutions constructed and analyzed by \refs{\LuninUU,\GiustoID,\GiustoIP,\GiustoKJ}.  However, a single
generic blob will have $J_1 - J_2$ small compared to $|J_1|$ and $|J_2|$.

To compute the other properties of such a blob,  we will simplify things
by taking $N=2 M +1$ to be odd, and assume that  $q_j = (-1)^{j+1}$.
Using the gauge invariance, we
can  take  all of  $k^I_i$ to be  positive numbers, and we will assume that
they have small variations about their mean value:
\eqn\kvariations{k^I_j  ~=~ k_0^I  \, (1 ~+~ \cO(1)) \,,}
where $k_0^I $ is defined in \ktilde.   The charges are given by:
\eqn\charges{\eqalign{Q_I &= - 2 \, C_{IJK }
\sum_j  q_j^{-1}\, (k^J_j- q_j N k_0^J )\, (k^K_j- q_j N k_0^K )  \cr
&=~ - 2 \, C_{IJK } \bigg( \sum_j q_j^{-1} k^J_j k^K_j - N k_0^J
\sum_j k^K_j -  N k_0^K \sum_j k^J_j + N^2 k_0^J k_0^K \sum_j q_j \bigg) \cr
&=~ 2\, C_{IJK } \Big( N^2 k^J k^K  - \sum_j k^J_j k^K_j q_j^{-1}\,   \Big)\cr
&\approx~ 2\, C_{IJK }  \big( N^2  +  \cO(1)\big )\, k_0^J k_0^K   }}
where we used \kvariations\ and the fact that $|q_i| =1$ only in the last step.
In the large $N$ limit,  for M theory on $T^6$ we have
\eqn\MtheoryQ{Q_1 \approx  4 N ^2 k^2 k^3 + \cO(1)\,, \quad Q_2 \approx
4 N^2 k^1 k^3 + \cO(1)\,, \quad Q_3 \approx 4 N^2 k^1 k^3 + \cO(1)\,.}

We can make a similar computation for the angular momenta:
\eqn\Jfin{\eqalign{J_R &~=~ \coeff{4}{3}\,  C_{IJK} \sum_j  q_j^{-2}\,
(k^I_j- q_j N k_0^I)\, (k^J_j- q_j N k_0^J )\, (k^K_j- q_j N k_0^K ) \cr
&~=~  \coeff{4}{3}\,  C_{IJK} \bigg( \sum_j   q_j^{-2}\,k^I_j k^J_j k^K_j
- 3  N k_0^I  \sum_j  q_j^{-1}\, k^J_j  k^K_j  \cr & \qquad \qquad \qquad
+ 3   N^2 k_0^I k_0^J \sum_j  k^K_j  -  N^3 k_0^I k_0^J  k_0^K
\sum_j q_j  \bigg) \cr
& ~ \approx~  \coeff{4}{3}\,  C_{IJK}
\big( N   - 3  N   + 3  N^3   - N^3  + \cO(N) \big) \, k_0^I k_0^J k_0^K\,,
}}
where we used the fact that, for a ``well behaved'' distribution of
positive $k_i^I$ with $|q_j| =1$,  one has:
\eqn\sums{  \sum_i q_i^{-1}  k^J_i  k^K_i ~=~
\sum_i q_i   k^J_i  k^K_i ~\approx~  k_0^J  k_0^K \,, \qquad
\sum_i  k^I_i k^J_i k^K_i ~\approx~ N k_0^I  k_0^J k_0^K \,.}
For M theory on $T^6$  we simply have:
\eqn\Japprox{J_R ~\approx~ 16 N^3 k^1 k^2 k^3 + O(N) ~.}

For large $N$ we therefore have, at leading order:
\eqn\bmpv{J_1^2 ~\approx~ J_2^2 ~\approx~ \coeff{1}{4}\, J_R^2 ~\approx~
 Q_1 Q_2 Q_3 \,,}
and so, in the large-$N$ limit, these microstates always correspond to a
maximally spinning BMPV black hole.  Indeed, we have
\eqn\deviation{{J_R^2 \over 4  Q_1 Q_2 Q_3} ~-~ 1\sim
O\left({1 \over N^2}\right)\,.}
Interestingly enough, the value of $J_R $ is slightly bigger than
$\sqrt{4 Q_1 Q_2 Q_3}$. However, this is not a problem because in the
classical limit this correction vanishes. Moreover, it is possible to argue
that a classical black hole with zero horizon area will receive  higher-order
curvature corrections, that usually increase the horizon area; hence a
zero-entropy configuration will have  $J_R$ slightly larger then the maximal
classically allowed value, by an amount that vanishes in the large $N$ (classical) limit.

\subsec{Zero-entropy black ring (supertube) microstates}

The next simplest configuration to consider is one in which one starts
with  the blob considered above and then moves a single GH point of charge
$+1$ out to a very large distance from the blob.  That is, one considers a blob
of total GH charge zero with a single very distant point of GH-charge $+1$.
Since one now has a strongly ``biased'' distribution of GH charges one
should now expect $J_1 - J_2 \ne 0$.

Again we will assume $N$ to be odd, and take the
a  GH charge distribution to be $q_j = (-1)^{j+1}$, with the distant
charge being the $N^{\rm th}$ GH charge.   We will also fix the gauge
by taking $k_N^I \equiv 0$.
The blob therefore has ${1 \over 2}(N-1)$ points of GH charge $\pm 1$ and
at large scales one might expect it to resemble the three-point
solution described above with $Q= {1 \over 2}(N-1)$.  We will show that
this is exactly what happens in the large-$N$ limit.

To have  the $N^{\rm th}$ GH charge far from the blob means that
 all the two-cycles, $\Delta_{j \, N}$ must support a very large flux
 compared to the fluxes on the $\Delta_{ij}$ for $i,j < N$.
 To achieve this we therefore take:
\eqn\ringblobvars{k^I_j  ~=~ a_0^I  \, (1 ~+~ \cO(1))  \,, \quad j=1,\dots,N -1 \,, \qquad
k^I_N  ~=~ - b_0^I \, N \,.}
where $b_0^I$ is independent of $j$ and
\eqn\avariations{a^I_j ~\equiv~k^I_j  ~=~ a_0^I  \, (1 ~+~ \cO(1)) \,,
\quad j=1,\dots,N -1 \,,  }
with
\eqn\aavg{a_0^I ~\equiv~{1 \over (N-1)} \, \sum_{j=1}^{N-1} a_j^I\,.}
We also assume that $a_0^I$ and $b_0^I$ are of the same order.
The fluxes of this configuration are then:
\eqn\blobfluxes{\Pi^{(I)}_{ij} ~=~ \bigg({ a_j^I \over q_j} ~-~  { a_i^I \over q_i}
\bigg) \,, \qquad \Pi^{(I)}_{i\,N} ~=~  - \Pi^{(I)}_{ N \, i} ~=~ -\bigg(
{ a_i^I  \over q_i} ~+~    N\, b_0^I \bigg) \,, \quad i,j =1, \dots, N-1\,.}
For this configuration one has:
\eqn\abvariations{\eqalign{k_0^I  ~=~ & {(N-1)\over N} \, a_0^I ~-~b_0^I \,,
\qquad \tilde k_N^I  ~=~  - (N-1) \, a_0^I   \,, \cr
\qquad  \tilde k^I_j  ~=~ & a_j^I ~+~ q_j \, (N\, b_0^I  - (N-1)\, a_0^I)\,, \quad
j=1,\dots, N-1 \,.}}
Motivated by the bubbling black ring of \BenaVA, define the physical parameters:
\eqn\dandf{d^I ~\equiv~  2\,(N-1)\, a_0^I \,, \qquad f^I ~\equiv~
(N-1)\, a_0^I - 2\,N\, b_0^I \,.}
Keeping only the terms of leading order in $N$
in \QIchg\ and \Jright, one finds:
\eqn\rcharges{Q_I ~=~ C_{IJK} d^J f^K \,, \qquad
J_1 + J_2  ~=~ \coeff{1}{2} \, C_{IJK}  (d^I d^J f^K + f^I f^J d^K) ~-~
 \coeff{1}{24} \, C_{IJK}  d^I d^J d^K\,.}

Since the $N^{\rm th}$ point is far from the blob,
we can take $r_{i N} \approx r_0$ and then the $N^{\rm th}$ bubble
equation reduces to:
\eqn\bubblerad{\coeff{1}{6}\, C_{IJK} \,\sum_{j=1}^{N-1}\,
\bigg({a_j^I \over q_j} + N\, b_0^I\bigg)\, \bigg({a_j^J \over q_j} + N\, b_0^J\bigg)\,
\bigg({a_j^K \over q_j} + N\, b_0^K\bigg)\, {q_j  \over r_0}~=~  (N-1)\, \sum_I \, a^I\,.}
To leading order in $N$ this means that the distance from the
blob to the $N^{\rm th}$ point, $r_0$, in the GH space is given by:
\eqn\ringrad{r_0 ~\approx~   \coeff{1}{2} \, N^2\, \bigg[\sum_I \, a^I\bigg]^{-1}
C_{IJK} \,   a_0^I \, b_0^J \, b_0^K ~=~  \coeff{1}{32} \,
\bigg[\sum_I \, d^I\bigg]^{-1}  C_{IJK} \,   d^I \, (2f^J - d^J) \, (2f^K - d^K) \,.}

Finally, considering  the dipoles, \dipoles, it is evident that to leading
order in $N$, $\vec D$ is dominated by the contribution coming from the
$N^{\rm th}$ point and that:
\eqn\ringJL{\eqalign{J_1 - J_2 ~=~ & 8\, | \vec D| ~=~  8\, N \, \bigg(
\sum_I \, a_0^I \bigg)\,  r_0 ~=~ 4 \, N^3 \, C_{IJK} \,   a_0^I \, b_0^J \, b_0^K
\cr  ~=~ & \coeff{1}{8} \,    C_{IJK} \,   d^I \, (2f^J - d^J) \, (2f^K - d^K) \,.}}

One can easily verify that these results perfectly match the properties
of the three-center bubbled supertube constructed in \BenaVA, and summarized
above.  Thus the blob considered here  has {\it exactly} the  same size,
angular momenta, charges and dipole charges as a zero-entropy black ring.

\subsec{Numerical investigation of bubbling solutions}

It is extremely instructive to solve the bubble equations
numerically for fairly large values of $N$ because one discovers some
interesting qualitative results.    Most particularly, the interior structures
of bubbled black holes and bubbled black rings is  very different, at least
when all the flux parameters are equal.  The black ring blob has zero total
GH charge, and the individual GH charges form tight, neutral clusters that are
fairly broadly spaced.   The black hole blob on the other hand has a net GH charge,
which prevents the formation of neutral clusters everywhere. In fact the
GH charges do form broadly-spaced tight, neutral clusters but only in the
outer parts of the blob.  The excess GH charge can be found in the
deep interior of the blob, where it is ``screened'' by the neutral clusters.

\goodbreak\midinsert
\vskip .2cm
\centerline{ {\epsfxsize 3.6in\epsfbox{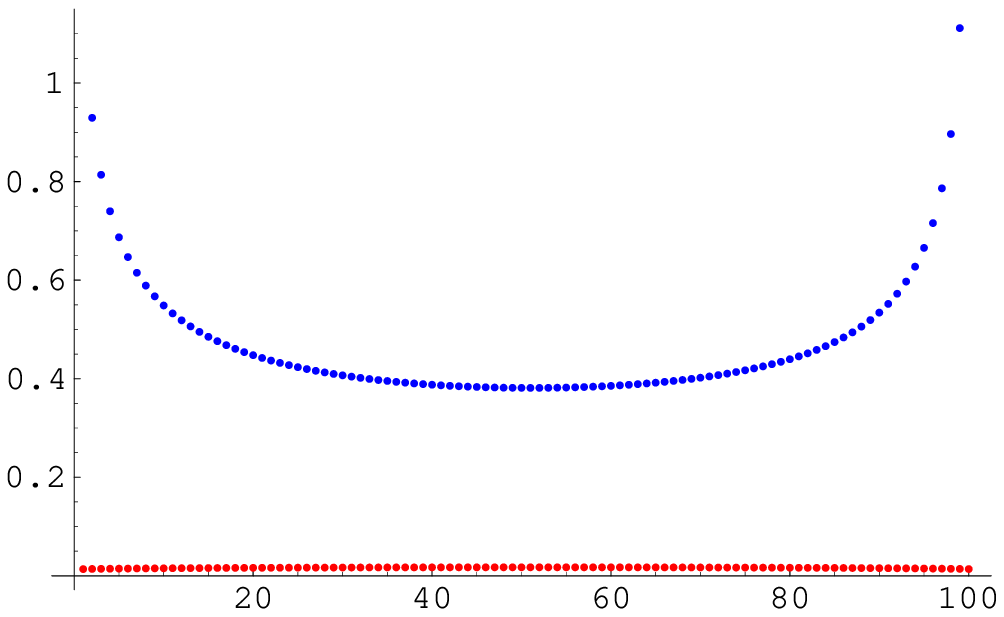}}}
\vskip -.25cm
\leftskip 2pc
\rightskip 2pc\noindent{\ninepoint\sl \baselineskip=8pt
{\bf Fig.~2}:  A ring distribution for $N = 201$ points in which
the $j^{\rm th}$ point has GH charge  $(-1)^{j+1}$.   There is a
fixed flux parameter, $k =1$,  on first $(N-1)$ points, and a flux
parameter $20 k$ on the
$N^{\rm th}$ point.  The width of the ring distribution (the distance from
first to $(N-1)^{\rm th}$ point)  is $49.09$. The distance from
$(N-1)^{\rm th}$ to $N^{th}$  point is $376.41$.   This graph shows distances
between successive GH points. The  lower curve  shows the spacing between
the $(2j-1)^{\rm th}$ point  and the $(2j)^{\rm th}$ point, $j=1, \dots, 100$,
forming $(+-)$ dipole pairs.  The upper curve   shows the spacing
between the $(2j)^{\rm th}$ point  and the $(2j+1)^{\rm th}$
point, $j=1, \dots, 99$, representing the separations of neighboring dipoles.
The ``far away'' point is not shown.
Note that the entire distribution shows a set  of $100$ close (+ -) dipoles.
All distances are measured using the flat $\IR^3$ part of the GH metric.}
\endinsert

We solved numerically the bubble equations for a very large number  of GH
centers lying on the same axis, with alternating charges $q_j = \pm 1$.
To get the ``ring blob'' we  took the flux parameters $k_i$ to be equal
on the first $(N-1)$ points and adjusted the flux parameter on the last point
so as to move it away from the blob.  We generically found the solution to
consist of tightly bound dipoles in the blob and that the spacing
between two neighboring dipoles was at least an order of magnitude larger that the
size of an individual dipole.    The ratio of the separation of neighboring dipoles
and the size of a dipole grows  larger as the $N^{\rm th}$ ``far away'' GH point
is moved further and further from the blob.  Figure 2 shows a
typical set of spacings between successive GH points.  There are two
branches to this graph:  The size of dipoles and the space between
neighboring dipoles.

\goodbreak\midinsert
\vskip .2cm
\centerline{ {\epsfxsize 3.6in\epsfbox{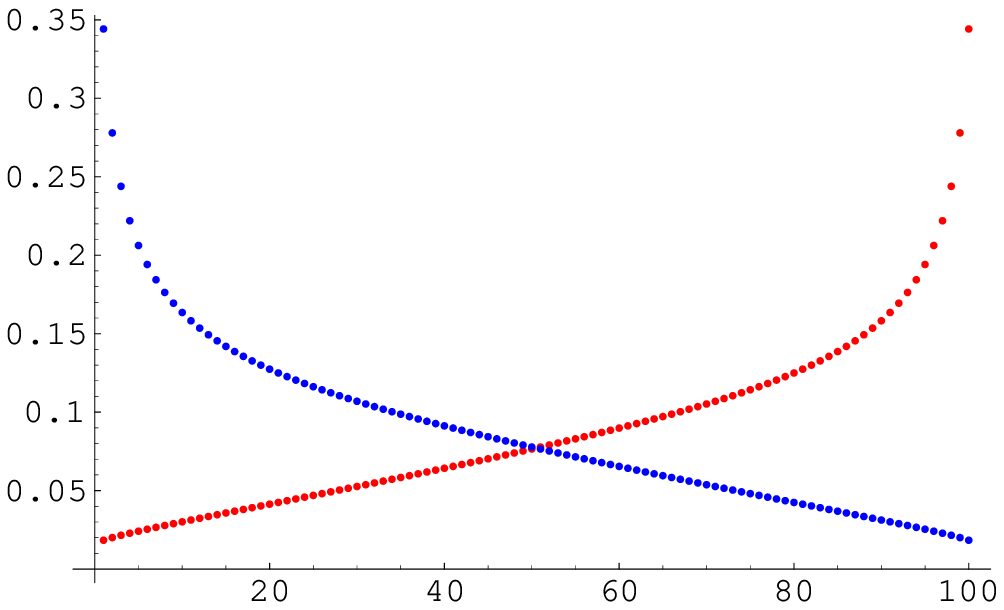}}}
\vskip -.25cm
\leftskip 2pc
\rightskip 2pc\noindent{\ninepoint\sl \baselineskip=8pt
{\bf Fig.~3}:  A BMPV blob with $N = 201$ points and all
flux parameters equal.  The $j^{\rm th}$ point has GH charge
$(-1)^{j+1}$.  This graph shows distances  between neighboring GH points.
The  curve from bottom-left to top-right shows the spacing between the $
(2j-1)^{\rm th}$ point and the $(2j)^{\rm th}$ point, a $(+-)$-pair, $j=1, \dots, 100$.
The  curve from top-left to  bottom-right shows the spacing between the
$(2j)^{\rm th}$ point   and the $(2j+1)^{\rm th}$ point, a $(-+)$-pair,
$j=1, \dots, 100$.  Note how the points are evenly spaced in the middle
and form  a more diffuse ``gas'' of close dipoles of opposite orientation
at the edges.  }
\endinsert

To get a BMPV blob we took all the flux parameters to be equal
and sought solutions where the GH charges alternated along
a symmetry axis in $\IR^3$.   All the solutions we found had a further
reflection symmetry in the distribution of GH charges, and this
symmetry guarantees that $\vec D \equiv 0$, and so $J_1 = J_2$.
 The distribution of spaces between GH points  is completely
 different from that of the ``ring blob.'' By construction, the outermost points
had GH charge $+1$ and we found that the charges formed close dipoles with their
nearest neighbors.  Thus the outer part of the distribution tries to
form $(+-)$ dipoles on the left and $(-+)$ dipoles on the right.  This
cannot be sustained in the interior and so the dipole length grows and
the charges become evenly-spaced towards the middle of the configuration.
The excess GH charge of the blob is thus in the middle and it tends to separate
the nearby dipoles.  Figure 3 shows a  typical set of spacings between
successive GH points in a BMPV blob. Again there are two branches,
one for the $(+-)$-pairs and another for the $(-+)$-pairs.  These two
branches cross in the center.

One important consequence of this is that a black-ring blob and a black-hole blob
are very different from each other, despite having similar flux parameters.
Moving a single GH point out of a large blob, might naively appear to be a small
perturbation; it can however result in a huge ``phase transition'' in the interior
structure of the blob.  In particular, moving a single $+1$ point from the edge
of the ``BMPV blob'' in Figure 3 should result in a transition to the distribution shown
in Figure 2.   We will see an example of this in Section 6.

\newsec{Mergers and microstates}

Having found some of the microstates corresponding to the maximally-spinning
BMPV black hole and to a supertube (zero-entropy black ring), we now revisit the merger
process of a black ring  and a black hole from the perspective of microstates.  We
begin by recalling some of the key results of \BenaZY.

\subsec{Classical mergers and their bubbled counterparts}

Consider the $U(1)$-invariant configurations of a black ring
and a black hole in which the black hole lies on the axis of the ring,  but
offset  ``vertically'' from the center.  These solutions were first obtained
and studied  in  \BenaZY.   In particular, for fixed
macroscopic charges on the ring and black hole, it was shown how
the embedding radius of the ring must vary as one varies the offset.
In this process, one component, $J_1$, of the angular momentum in the ring
plane is  conserved, while the other component, $J_2$, changes.
The merger of the black ring with the black hole occurs precisely
when $J_L  \equiv J_1 -J_2 =0$.

Generically  such a merger is thermodynamically irreversible
and results in a BMPV black hole ($J_1 =J_2$) with a total horizon
area that exceeds the sum of the areas of the two original horizons.
There is, however, one possible way to make a reversible merger:
The initial black hole and black ring have to have vanishing horizon area and
these could merge to yield  a larger, maximally-spinning (zero-horizon-area)
black hole provided  that two conditions are met. First, the ring must merge
with the black hole at the equator of the black hole, so that at the point of merger
the solution has a $U(1) \times U(1)$ invariance.  Secondly, the charge vectors
of the ring and of the black hole must be parallel.  Any other merger will
result in an entropy increase.

One can also study the merger by considering a $U(1) \times
U(1)$ invariant solution describing a black ring with a black hole in the center \refs{\GauntlettWH,\BenaDE,\GauntlettQY}.
As the ring is made smaller and smaller by, for example, decreasing its
angular momentum, it eventually merges into the black hole.  At the point of
merger, this solution is identical to the merger described above with the black ring
grazing the black-hole horizon.  Hence this $U(1) \times U(1)$ invariant solution
can be used to study mergers where the black ring grazes the black hole
horizon at the point of merger; all the reversible mergers and some of the
irreversible mergers belong to this class.

In the previous section we have seen how to create bubbled solutions
corresponding to zero-entropy black rings and maximally-spinning black holes.
The generic bubbled solutions with GH base have a $U(1)$ symmetry corresponding
to $J_R \equiv  J_1 + J_2$ and if the GH points all lie on an axis then the solution is
$U(1) \times U(1)$ invariant.    Conversely \GibbonsSP,
all $U(1) \times U(1)$ invariant, four-dimensional hyper-K\"ahler metrics must
be of GH form.  We can therefore study the merger of bubbled
microstates by constructing  $U(1) \times U(1)$ invariant bubbling solutions
describing a black ring with a black hole in the center. By decreasing some
of the flux parameters of the solution one can decrease the radius of the bubbling
black ring and merge it into the bubbling black hole to create a larger bubbling black hole.

In this section we consider a bubbling black hole with a very large number
of GH centers, sitting at the center of the simplest bubbled supertube \BenaVA,
generated by a pair of GH points\foot{Of course it is  straightforward to generalize
our analysis to the situation where both the supertube and the black hole have a
large number of GH centers. However, the analysis is simpler and the numerical stability
is better for mergers in which the supertube is composed of only two points,
and we have therefore focused on this.}.  We expect two different
classes of merger solution depending upon whether the flux parameters
on the bubbled black hole and bubbled black ring are parallel or not.  These correspond
to reversible and irreversible mergers respectively.
We will examine both situations in some detail, particularly when all the
GH points lie on an axis and in which the solution has a $U(1) \times U(1)$
symmetry, and indeed find that reversible and irreversible mergers
of microstates have very different physics. First, however, we will establish some
general results about the charges and angular momenta of the bubbled solutions
that describe a bubbled black ring of two GH centers with a BMPV black-hole
blob at the center.

\subsec{Some exact results}

We begin by seeing what may be deduced with no approximations whatsoever.
Our purpose here is to separate all the algebraic formulae for charges
and angular momenta into those associated with the black hole foam  and those
associated with the bubbled supertube.   We will consider a  system of $N$ GH points
in which the first  $N-2$  points will be considered to be a blob and the
last two points will have $q_{N-1} = -Q$  and $q_{N} = Q$.  The latter
two points can then be used to define a bubbled black ring.

Let $\hat k_0^I$ denote the average of the flux parameters
over the first $(N-2)$ points:
\eqn\kIavg{\hat k_0^I ~\equiv~{1 \over (N-2)} \, \sum_{j=1}^{N-2} \, k_j^I \,,}
and introduce $k$-charges that have a vanishing average  over the
first $(N-2)$ points:
\eqn\khat{\hat  k^I_j ~\equiv~ k^I_j ~-~ (N-2) \,  q_j  \,  \hat k_0^I  \,,
\qquad j=1,\dots,N-2 \,.}
We also parameterize the last two $k^I$-charges in exactly the same
manner as for the bubbled supertube (see \dfring):
\eqn\ringchgs{d^I ~\equiv~ 2 \, \big(  k_{N-1}^I + k_{N}^I \big) \,, \qquad
f^I ~\equiv~ 2 \,(N-2) \,   \hat k_0^I +  \big(1 + \coeff{1}{Q}\big)\, k_{N-1}^I  +
 \big(1 - \coeff{1}{Q}\big)\, k_{N}^I  \,.}
One can easily show that the charge \QIchg\ decomposes into
\eqn\chgdecomp{Q_I ~=~ \widehat Q_I  ~+~C_{IJK}\, d^J\, f^K \,,}
where
\eqn\QIhat{\widehat Q_I ~\equiv~ - 2\, C_{IJK} \, \sum_{j=1}^{N-2}  q_j^{-1} \,
\hat  k^J_j \, \hat  k^K_j  \,,}
The $\widehat Q_I$ are simply the charges of the black-hole blob, made of the
first $(N-2)$ points.  The second term in \chgdecomp\
is exactly the expression, \simpringchg, for the charges of
a bubbled supertube with GH centers of charges $+1$, $-Q$ and $Q$ and
$k$-charges $(N-2) \hat k_0^I$, $k_{N-1}^I$ and $k_{N}^I$, respectively.
Thus the charge of this configuration decomposes exactly as if
it were a black-hole blob of $(N-2)$ centers and a bubbled supertube.

There is a similar result for the angular momentum, $J_R$.  One
can easily show that:
\eqn\Rangmomdecomp{J_R ~=~ \widehat J_R  ~+~  d^I \,
\widehat Q_I~+~j_R\,,}
where
\eqn\JRhat{\hat J_R  ~\equiv~  {4 \over 3}\, \, C_{IJK} \,
\sum_{j=1}^{N-2} q_j^{-2} \,  \hat  k^I_j \, \hat  k^J_j \,  \hat  k^K_j  \,,}
and
\eqn\JringR{j_R  ~\equiv~ \coeff{1}{2} \, C_{IJK}\, \big(f^I f^J d^K ~+~
f^I  d^J  d^K\big)  ~-~ \coeff{1}{24} \,(1-Q^{-2})\, C_{IJK}\,  d^I d^J d^K \,.}
The term, $\hat J_R$, is simply the right-handed
angular momentum of the black-hole blob made from $N-2$ points.
The ``ring'' contribution to the angular momentum, $j_R$, agrees precisely
with $J_1+ J_2$ given by \jonering\ and \jtworing\ for an isolated bubbled supertube.
The cross term, $d^I \widehat Q_I$ is represents the interaction of the flux of
the bubbled ring and the charge of the black-hole blob.  This interaction term is
exactly the same as that found in  \refs{\BenaDE, \GauntlettQY, \BenaZY} for a
concentric  black hole and black ring.

Thus, as far as the charges and $J_R$ are concerned, the complete
system is behaving as though it were a black-hole blob of $(N-2)$
points interacting with a bubbled supertube defined by the points
with GH charges $\pm Q$ and a single point with GH charge $+1$
replacing the black-hole blob. Note that no approximations were made
in the foregoing computations, and the results are true independent
of the locations of the GH charges.

To make further progress we need to make some assumptions about
the configuration of the points.    Suppose, for the moment, that all the GH
charges lie on the $z$-axis at points $z_i$ with $z_i < z_{i+1}$.
In particular, the GH charges, $-Q$ and $+ Q$,  are located at $z_{N-1}$ and
$z_{N}$ respectively.

With this ordering of the GH points, the expression for $\vec J_L$ collapses to:
\eqn\linJL{\eqalign{ J_L ~=~  \coeff{4}{3}\, C_{IJK} \, \sum_{1 \le i < j \le N}
\,  q_i \, q_j \,  \Pi^{(I)}_{ij} \,   \Pi^{(J)}_{ij} \,  \Pi^{(K)}_{ij}   \,.}}
This expression can then be separated, just as we did for $J_R$, into a black-hole
blob component, a ring component, and interaction cross-terms.  To that
end, define the left-handed angular momentum of the blob to be:
\eqn\linJLh{\widehat J_L ~=~  \coeff{4}{3}\, C_{IJK} \, \sum_{1 \le
i < j \le N-2} \,  q_i \, q_j \,  \Pi^{(I)}_{ij} \,   \Pi^{(J)}_{ij}
\,  \Pi^{(K)}_{ij}   \,.}
Note that
$$
\Pi^{(I)}_{ij} ~\equiv~  \bigg( {k_j^I \over q_j} ~-~
{k_i^I \over q_i} \bigg)  ~=~  \bigg( {\hat k_j^I \over q_j} ~-~
{\hat k_i^I \over q_i} \bigg)  \,, \qquad  1 \le i, j \le N-2 \,,
$$
and so this only depends upon the fluxes in the blob.

The remaining terms  in \linJL\ may then be written in terms of $\hat k_j^I$,
$d^I$ and $f^I$ defined in \khat\ and \ringchgs.  In particular, there are terms
that depend only upon $d^I$ and $f^I$, and then there are terms that are
linear, quadratic and cubic in $\hat k_j^I$ (and depend upon $d^I$ and $f^I$).
The linear terms vanish  because $\sum_{j=1}^{N-2}\hat k_j^I \equiv 0$, the quadratic terms
assemble into $\widehat Q_I$  of \QIhat\ and the cubic terms assemble
into $\hat J_R$ of  \JRhat.    The terms proportional to \JRhat\ cancel between
the terms with $j=N-1$ and $j=N$, and one is left with:
\eqn\Langmomdecomp{J_L ~=~ \widehat J_L  ~-~  d^I \,
\widehat Q_I~+~j_L\,,}
where $j_L$ is precisely the angular momentum, $J_T$, of the tube:
\eqn\JringL{j_L ~\equiv~ \coeff{1}{2} \, C_{IJK}\, \big(d^I f^J f^K ~-~
f^I d^J d^K \big)  ~+~ \Big( {3 \,Q^2 - 4\,Q +1 \over 24\, Q^2} \Big) \,
C_{IJK}\,  d^I d^J d^K \,.}
Observe that \JringR\ and \JringL\  are exactly the angular momenta
of the simplest bubbled supertube, \jonering\ and \jtworing.  Again we see
the cross-term from the interaction of the ring dipoles and the
electric charge of the blob.  Indeed, combining \Rangmomdecomp\
and \Langmomdecomp, we obtain:
\eqn\angmomcomp{J_1 ~=~ \widehat J_1   ~+~j_1\,, \qquad
J_2 ~=~ \widehat J_2   ~+~ j_2 ~+~  d^I \,  \widehat Q_I  \,,}
which is exactly how the angular momenta of the classical ring-hole solution
decompose \BenaZY.   In particular, the term coming from the interaction of the
ring dipole moment to the black hole charge only contributes to $J_2$.

The results obtained above are independent of whether the blob of
$N-2$ points is a BMPV black-hole blob, or a more generic configuration.
However, to study mergers, we will explore from now on configurations in which the blob is a black-hole
microstate, with $\widehat J_L ~=~ 0$.  The end result of the merger process is
also a BMPV black hole microstate, and so $J_L ~=~  0$. Therefore, the {\it exact}
merger condition is simply:
\eqn\exactMcond{\eqalign{\Omega &~\equiv~ \coeff{1}{2} \, C_{IJK}\,
\big(d^I f^J f^K ~-~  f^I d^J d^K \big)  ~+~ \Big( {3 \,Q^2 - 4\,Q +
1 \over 24\, Q^2} \Big) \,  C_{IJK}\,  d^I d^J d^K ~-~  d^I \,  \widehat Q_I\cr
&~=~0 \,.}}
Using \Jsimptube, this may be written:
\eqn\ClassMerger{ J_T ~-~  d^I \,  \widehat Q_I ~=~0 \,,}
which is precisely the condition obtained in \BenaZY\ for a classical
black ring to merge with a black hole at its equator.

One should note that the argument that led
 to the expressions \Langmomdecomp\ and \JringL,  and the
 exact merger  condition, \exactMcond, apply far more generally.  In particular
we only needed the fact that the unit vectors, $\hat y_{ij}$, defined
 \unitvecs, are all parallel for $j = N-1$ and $j=N$.   This is
 approximately true in may contexts, and most particularly if the
 line between the  $(N-1)^{\rm th}$ and  $N^{\rm th}$ points runs
 through the blob and the width of the  blob is small compared
 to the distance to these two points.

One should also not be surprised by the generality of the result in equation \JringL.
 The angular momentum, $J_T$, is an  intrinsic property of a black ring, and hence
for a zero-entropy black ring, $J_T$ can only depend on the $d$'s
and $f$'s, and cannot depend on the black hole charges (that is, the $\hat k_j^I$).
Therefore, we could have obtained \JringL\ by simply setting the black hole charge
to zero, and then reading off $J_T$ from the simplest bubbling black ring solution \BenaVA\ discussed in sub-section 2.3. Hence, one should think
about the expression of  $J_T$ in \Jsimptube\ as a universal relation between
intrinsic properties of the bubbled ring: $J_T, d^I$ and $f^I$.

\subsec{Some simple approximations}

We now return to a general distribution of GH points, but we will assume
that the two ``exceptional points'' (the $(N-1)^{\rm th}$ and $N^{\rm th}$ points)
are close  together but very far from the black-hole blob of the remaining $(N-2)$
points.   Set up coordinates in the geometric center of the black-hole blob, {\it i.e.}
choose the origin so that
\eqn\geomcent{\sum_{i=1}^{N-2} \, \vec r_i ~=~ 0 \,.}
Let $r_0  \equiv |\vec r_{N-1}|$ be the distance from the geometric center
of the blob to the first exceptional point, and let $\hat r_0$ be the unit
vector in that direction.  The vector, $\vec \Delta \equiv \vec r_{N } - \vec r_{N-1}$,
defines the width of the ring.   We will assume  that the
magnitudes $\Delta \equiv  |\vec \Delta|$  and $ r_j \equiv  |\vec  r_j|$ are small
compared to $r_0$.  We will also need the first terms of the multipole expansions:
\eqn\dipoleexp{\eqalign{{1\over |\vec r_{N-1}-\vec r_j |} ~=~ &
{1 \over r_0} ~+~  { \vec r_j \cdot \hat r_0  \over r_0^2}  ~+~ \dots  \cr
{1\over | \vec r_N  -\vec r_j |} ~=~ &{1 \over r_0} ~+~
{(\vec r_j - \vec \Delta)\cdot \hat r_0  \over r_0^2}  ~+~ \dots \,.}}
For simplicity, we will also assume that the two  ``exceptional points''
are co-linear with the origin so that
\eqn\rNval{r_N ~\equiv~  | \vec r_{N}  | ~=~ r_0 + \Delta \,.}

The last two bubble equations are then:
\eqn\bubbleone{ { \gamma \over  \Delta} ~-~  \sum_{j=1}^{N-2}\,
{q_j\, \alpha_j  \over  |\vec r_N - \vec r_j|}   ~=~  \sum_I \,
\big(N \,Q\, k^I_0 - k_{N}^I\big) \,,}
\eqn\bubbletwo{ - { \gamma \over  \Delta} ~+~  \sum_{j=1}^{N-2}\,
{q_j\, \beta_j   \over |\vec r_{N-1} -  \vec r_j|}   ~=~ - \sum_I \,
\big(N \,Q\, k^I_0  + k_{N-1}^I\big)\, }
where $k^I_0$ is given in \ktilde\ and
\eqn\aldefn{\eqalign{\alpha_j ~\equiv~& \coeff{1}{6} \,Q\, C_{IJK}
\, \Pi_{j\,N}^{(I)} \,  \Pi_{j\,N}^{(J)}\,  \Pi_{j\,N}^{(K)} \cr
~=~&   \coeff{1}{6} \,Q\,  C_{IJK} \,
\bigg({k_{N}^I \over Q} - {k_j^I \over q_j} \bigg)\, \bigg({k_{N}^J \over Q} -
{k_j^J \over q_j} \bigg)\,\bigg({k_{N}^K \over Q} - {k_j^K \over q_j} \bigg) \,,}}
\eqn\bedefn{\eqalign{\beta_j ~\equiv~& \coeff{1}{6}  \,Q\, C_{IJK}
\, \Pi_{j\,(N-1)}^{(I)} \,  \Pi_{j\,(N-1)}^{(J)}\,  \Pi_{j\,(N-1)}^{(K)} \cr
~=~& - \coeff{1}{6} \,Q\,  C_{IJK} \,
\bigg({k_{N-1}^I \over Q} + {k_j^I \over q_j} \bigg)\, \bigg({k_{N-1}^J \over Q} +
{k_j^J \over q_j} \bigg)\,\bigg({k_{N-1}^K \over Q} + {k_j^K \over q_j} \bigg)  \,,}}
\eqn\gadefn{\gamma ~\equiv~  \coeff{1}{6}\, Q^2 \, C_{IJK}\,
\Pi_{ (N-1)\,N}^{(I)} \,  \Pi_{ (N-1)\,N}^{(J)} \, \Pi_{ (N-1)\,N}^{(K)}
~=~  \coeff{1}{48}\,  Q^{-1} \,  C_{IJK} \, d^I\, d^J\,d^K  \,.}
It is also convenient to introduce
\eqn\ABzero{\alpha_0 ~\equiv~ \sum_{j=1}^{N-2} \,q_j \, \alpha_j\,, \qquad
\beta_0 ~\equiv~ \sum_{j=1}^{N-2} \,q_j \, \beta_j\,.}

If one adds \bubbleone\ and \bubbletwo\  then the terms involving $\gamma$
cancel and using  \dipoleexp\ one then obtains:
\eqn\expbubr{
\sum_{j=1}^{N-2}\,q_j\,  \bigg[\, \alpha_j \, \bigg({1 \over r_0}  +  {(\vec r_j - \vec \Delta)
\cdot \hat r_0 \over r_0^2}  \bigg)    ~-~ \beta_j  \, \bigg({1 \over r_0} +
 { \vec r_j \cdot \hat r_0  \over r_0^2} \bigg)\,\bigg] ~=~
 \coeff{1}{2} \,\sum_I \,  d^I \,.}
One now needs to perform the expansions with some care.   Introduce the flux
vector:
\eqn\Xdefn{X^I ~\equiv~  2\,f^I - d^I - 4\,(N-2)\,\hat k_0^I \,, }
and note that the fluxes between the blob and ring points are given by:
\eqn\ringholeflux{\Pi^{(I)}_{j \,(N-1)} ~=~ -\coeff{1}{4}\, \big[\,
X^I + Q^{-1}\, d^I  + 4\, q_j^{-1} \, k_j^I \,\big]\,, \qquad
\Pi^{(I)}_{j \,N} ~=~ -\coeff{1}{4}\, \big[\, X^I - Q^{-1}\, d^I  +
4\, q_j^{-1} \, k_j^I \,\big] \,.}
In particular, the difference of these fluxes is simply the flux through the two-cycle
running between the two ring points:
\eqn\fluxdiff{\Pi^{(I)}_{j \,N}~-~ \Pi^{(I)}_{j \,(N-1)} ~=~   {d^I  \over 2\, Q}  ~=~
\Pi^{(I)}_{(N-1)\, N} \,.}
For the ring to  be far from the black hole, the fluxes $\Pi^{(I)}_{j \,(N-1)}$ and
$\Pi^{(I)}_{j \,N}$ must be large.  For the ring to be thin ($\Delta \ll  r_0$), these fluxes
must be of similar order, or $\Pi^{(I)}_{(N-1)\, N}$ should be small.  Hence
we are assuming that $ {d^I  \over 2 Q} $ is small compared to $X^I$.
We are also going to want the black hole and the black ring to have similar
charges and angular momenta, $J_R$,  and one of the ways of achieving this
is to make $f^I$, $d^I$ and  $N \hat k_0^I$ of roughly the same order.

Given this,  the leading order terms in \expbubr\ become:
\eqn\redexpbubr{
\sum_{j=1}^{N-2}\, q_j\, \bigg[\, {(\alpha_j - \beta_j) \over r_0} ~-~ \alpha_j \,
{\Delta    \over r_0^2} \,\bigg] ~=~  \coeff{1}{2} \,\sum_I \,  d^I \,.}
One can then determine the ring width, $\Delta$, using \bubbleone\ or \bubbletwo.
In particular, when the ring width is small while the ring radius is large, the left-hand
side of each of these equations is the difference of two very large numbers of similar
magnitude.  To leading order we may therefore neglect the right-hand sides and
use the leading monopole term to obtain:
\eqn\deltares{\beta_0 \,{\Delta \over r_0} ~\approx~
\alpha_0 \,{\Delta \over r_0} ~=~  \bigg[\, \sum_{j=1}^{N-2}\, q_j\,
\alpha_j  \,\bigg] \,  {\Delta \over r_0}~\approx~\gamma    \,, }
and hence \expbubr\ becomes:
\eqn\simpexpbubr{
- \gamma   ~+~ \sum_{j=1}^{N-2}\, q_j\, (\alpha_j - \beta_j)    ~\approx~
 \bigg[\,  \coeff{1}{2} \,\sum_I \,  d^I\, \bigg] \, r_0 \,.}
Using the explicit expressions for $\alpha_j, \beta_j$ and $\gamma$, one
then finds:
\eqn\newringrad{\eqalign{r_0 ~\approx~  \bigg[4\, \sum_I \, d^I\bigg]^{-1}
 \bigg[ \, \coeff{1}{2} \, C_{IJK} \,      (d^I f^J  f^K & -   f^I d^J d^K)  \cr
 & ~+~ \bigg({3\,Q^2 -4\,Q +1 \over 24\, Q^2}\bigg) \,
C_{IJK} \, d^I d^J d^K  ~-~ d^I \widehat Q_I \, \bigg] \,.}}
This is exactly the same as the formula for the tube radius that one obtains
from \JTRreln\ and \Jsimptube.  Note also that we have:
\eqn\rradangmom{\eqalign{r_0 ~\approx~  \Big[4\, \sum_I \, d^I\Big]^{-1}
 \big[ \, j_L ~-~ d^I \widehat Q_I \, \big] \,,}}
where $j_L$ the angular momentum of the supertube  \Langmomdecomp.
In making the comparison to the results of \BenaZY, recall that for a black ring
with a black hole  exactly in the center, the embedding radius in
the standard, flat $\IR^4$ metric is given by:
\eqn\embedR{   R^2 ~=~{ l_p^6  \over L^4 }\,
\,\Big[\sum d^I \Big]^{-1} \,  \Big(J_T ~-~ d^I \widehat Q_I  \Big) \,.}
The transformation between a flat $\IR^4$ and the GH metric
with $V ={1 \over r}$ involves setting  $r = {1 \over 4} \rho^2$,   and this
leads to the relation  $R^2 = 4 R_T$.  We therefore have complete consistency
with the classical merger result.

Note that the classical merger condition is simply $r_0 \to 0$, which is, of course,
very natural.  This might, at first, seem to fall outside the validity of our
approximation, however we will see in the next section that for irreversible mergers
one does indeed maintain $\Delta, r_j \ll r_0$ in the limit  $r_0 \to 0$.
Reversible mergers cannot however be described in this approximation,
and have to be analyzed numerically. This is the subject of section 6.

\newsec{Irreversible mergers and scaling solutions}

We now show that an irreversible merger occurs in such a manner
that the ring radius, $r_0$, the ring width, $\Delta$, and a typical separation
of points within the black-hole blob all limit to zero while their ratios all limit to
finite values.  We will call these {\it scaling solutions}, or {\it scaling mergers}. As the
merger progresses, the throat of the solution becomes deeper and deeper, and
corresponding redshift becomes larger and larger. The high-redshift ``deep microstates" that result are microstates of a BPS back hole with classically large horizon area.

\subsec{Merging a ring and a black hole}

We use the solution discussed in the previous section, and we decrease the radius of
the bubbled ring, $r_0$  by decreasing some of its flux parameters. We take all the
flux parameters of the  $(N-2)$ points in the blob to be parallel:
\eqn\kpara{k_j^I ~=~ \hat k_0^I  ~=~ k^I\,, \qquad j~=~1, \dots, N-2
\,,}
Further assume that all the GH charges in the black-hole blob obey
$q_j = (-1)^{j+1}$, $j=1,\dots,N-2$.  We therefore have
\eqn\QJhat{\widehat Q_I ~=~ 2\,(N-1)(N-3)\, C_{IJK}\, k^J k^K \,, \qquad
\widehat J_R~=~ \coeff{8}{3}\,(N-1)(N-2)(N-3)\, C_{IJK}\, k^I k^J k^K  \,.}

Define:
\eqn\muidefn{\mu_i ~\equiv~  \coeff{1}{6}\, (N-2 - q_i)^{-1} \,
C_{IJK} \, \sum_{{\scriptstyle j=1} \atop {\scriptstyle j \ne
i}}^{N-2} \,  \,  \Pi^{(I)}_{ij} \,   \Pi^{(J)}_{ij} \,
\Pi^{(K)}_{ij} \ {q_j \over r_{ij} }  \,, }
then the bubble equations for this blob in isolation ({\it i.e.} with
no addition bubbles, black holes or rings) are simply:
\eqn\bhbubbles{\mu_i ~=~   \sum_{I=1}^3 \, k^I \,,}
More generally,  in any solution satisfying \kpara, if one finds a blob in which
the $\mu_i$ are all equal to the same constant, $\mu_0$, then the
GH points in the blob must all be arranged in the same way
as an isolated black hole, but with all the positions scaled by
$\mu_0^{-1}  \big(\sum_{I=1}^3 \, k^I \big)$.

Now  consider the full set of $N$ points with $\Delta, r_j \ll r_0$.  In the
previous sub-section we solved the last two bubble equations and
determined $\Delta$ and $r_0$ in terms of the flux parameters.
The remaining bubble equations are then:
\eqn\newbubblei{ (N-2 - q_i)\, \mu_i  ~+~ {\alpha_i \over | \vec r_N
- \vec r_i |} ~-~ {\beta_i \over | \vec r_{(N-1)} -\vec r_i |} ~=~
\sum_{I=1}^3 \, \big((N-2 - q_i)\,k^I  + {d^I \over 2} \big) \,, }
for $i =1,\dots, N-2$.  Once again we use the multipole expansion in these equations:
\eqn\ithmultipole{ (N-2 - q_i)\, \mu_i  ~+~  {(\alpha_i -\beta_i)  \over r_0}  ~-~
 {\alpha_i \,  \Delta  \over r_0^2} ~=~
\sum_{I=1}^3 \, \big((N-2 - q_i)\,k^I  + {d^I \over 2} \big) \,, }

It is elementary to show that:
\eqn\ABdiff{ \alpha_i - \beta_i ~=~ \coeff{1}{8}\, (j_L - d^I \,
\widehat Q_I ) ~+~  \gamma  ~-~
 \coeff{1}{8}\,  (N-2 - q_i)\, C_{IJK} \,  d^I\, k^J\, X^K \,,}
where $X^I $ is defined in \Xdefn.
If one now uses this identity, along with \deltares\ and \rradangmom\ in
\ithmultipole\ one obtains:
\eqn\newith{ \eqalign{(N-2 - q_i)\, \mu_i  ~-~  {1  \over  r_0} \,C_{IJK} \,\Big[
\coeff{1}{8}\,  (N-2 - q_i)\,  d^I\, k^J\, X^K & ~-~  \Big(1-{\alpha_i \over \alpha_0}\Big)\,
\gamma \Big]  \cr & ~\approx~  (N-2 - q_i)\, \sum_{I=1}^3 \, k^I   \,.}}
Finally, note that:
\eqn\AzeroAi{\alpha_0 -  \alpha_i ~=~ Q \,(N-2 - q_i)\,C_{IJK} \, \big[ \coeff{1}{32}
(X^I - \coeff{1}{Q} d^I)\, (X^J - \coeff{1}{Q} d^J)\,k^K ~+~ \coeff{1}{6}\,
k^I\, k^J\,k^K \big]\,,}
which means that all the terms in \newith\ have a factor of $(N-2 - q_i)$.
Hence the bubble equations \newbubblei\ reduce to:
\eqn\unieqns{ \eqalign{\mu_i ~\approx~&
\Big(\sum_{I=1}^3 \, k^I \Big)  ~+~ {1  \over  r_0} \,C_{IJK} \,
\Big[ \coeff{1}{8}\,  d^I\, k^J\, X^K  \cr & \qquad \qquad ~-~
 \alpha_0^{-1}\,Q\, \gamma\, \big( \coeff{1}{32}\,
(X^I - \coeff{1}{Q} d^I)\, (X^J - \coeff{1}{Q} d^J)\,k^K ~+~ \coeff{1}{6}\,
k^I\, k^J\,k^K \big) \,   \Big]    \cr  ~\approx~&
\Big(\sum_{I=1}^3 \, k^I \Big)  ~+~ {1  \over  r_0} \,C_{IJK} \,
\Big[ \coeff{1}{8}\,  d^I\, k^J\, X^K  ~-~ \alpha_0^{-1}\,Q\, \gamma\, \big(
\coeff{1}{32}\, X^I  \,  X^J  \,k^K ~+~ \coeff{1}{6}\, k^I\, k^J\,k^K \big) \,   \Big]   \,,}}
where we have used the assumption that $X^I$ is large compared to $Q^{-1} d^I$.

Observe that the right-hand side of \unieqns\ is independent of
$i$, which means that the first $(N-2)$ GH points satisfy a scaled
version of  the equations \bhbubbles\ for a isolated, bubbled black hole.
Indeed, if $\vec r_i^{BH}$ are the positions of a set of GH points satisfying
\bhbubbles\ then we can solve \unieqns\ by scaling the black hole solution,
$\vec r_i = \lambda^{-1} \vec r_i^{BH}$, where the scale factor is given by:
\eqn\scalefac{ \lambda ~\approx~
1~+~  {1  \over  r_0} \,\Big(\sum_{I=1}^3 \, k^I \Big)^{-1}\,C_{IJK} \,
\Big[ \coeff{1}{8}\,  d^I\, k^J\, X^K  ~-~ \alpha_0^{-1}\,Q\, \gamma\, \big(
\coeff{1}{32}\, X^I  \,  X^J  \,k^K ~+~ \coeff{1}{6}\, k^I\, k^J\,k^K \big) \,   \Big]  \,.}

Notice that as one approaches the critical ``merger'' value, at which
$\Omega = j_L - d^I \, \widehat Q_I  =0$, \scalefac\ implies that the distance,
$r_0$, must  also scale as $\lambda^{-1}$.  Therefore the merger process will
typically involve sending $r_0 \to 0$ while respecting the  assumptions made
in our approximations ($\Delta, r_i \ll r_0$).  The result will be
a ``scaling solution'' in which all distances in the GH base are vanishing
while preserving their relative sizes.

We have indeed verified this picture of the generic merger process by making
quite a number of numerical computations, one of which we will present in the next subsection.  We tracked one merger through a range
where the scale factor, $\lambda$, varied from about $4$ to well over $600$.
We have also checked that this scaling behavior is not an artefact of
axial symmetry.  We performed several numerical simulations in which
the GH points of the black-hole blob were arranged along a symmetry
axis but the bubbled ring approached the black-hole blob at various
angles to this axis.  We found that the scaling behavior was essentially
unmodified by varying the angle of approach.

An important exception to the foregoing analysis arises when the term proportional
to $r_0^{-1}$ in \unieqns\ vanishes to leading order.   In particular, this happens
if we violate one of the assumptions of our analysis, namely, if  one has:
\eqn\XIzero{X^I ~\equiv~  2\,f^I - d^I - 4\,(N-2)\,k^I ~\approx~0\,,}
to leading order in $Q^{-1} d^I$.   If $X^I$ vanishes one can see that, to leading
order, the merger condition is satisfied:
\eqn\Omform{ \Omega ~\equiv~ j_L - d^I \, \widehat Q_I ~=~
\coeff{1}{8}  \, C_{IJK} \, d^I\, \big[X^J X^K + 8\,(N-2)\, k^J X^K
- \coeff{1}{3}\, Q^{-2}\,(4\,Q -1) \, d^J d^K + 16  \,k^J k^K\big]
~\approx~0 \,,}
and so one must have $r_0 \to 0$.  However, the foregoing analysis
is no longer valid, and so the merger will not necessarily result in a
scaling solution.

An important example of this occurs when  $k^I$, $d^I$ and $f^I$ are all parallel:
\eqn\kdfpara{k^I ~=~ k\, u^I\,, \qquad d^I ~=~ d\, u^I\,, \qquad f^I ~=~ f\, u^I\,,}
for some fixed $u^I$.  Then the merger condition \Omform\ is satisfied
to leading order, only when $X \equiv (2\, f - d -4\, (N-2)\,k)$ vanishes.
We will discuss examples of this in the next section, where we will see
that the merger process does not involve scaling and the GH points of
the ring move very close to the GH points of the black hole.

For non-parallel fluxes it is possible to satisfy the merger condition, \Omform,
while keeping $X^I$ large, and the result is a scaling solution.

Even if it looks like irreversible mergers progress until the final size on the
base vanishes, this is an artifact of working in a classical limit an ignoring the
quantization of the fluxes. After taking this into account we can see from
\unieqns\ that $r_0$ cannot be taken continuously to zero because the $d^I, f^I, X^I$
and $k^I$ are integers of half-integers.
Hence, the final result of an
irreversible merger is a microstate of a  high, but finite, redshift and  whose throat
only becomes infinite in the classical limit.

In order to find the maximum depth of the throat, one has to find the smallest
allowed value for the size of the ensemble of GH points. During the
irreversible merger all the distances scale, the size of the ensemble of points will
be approximately equal to the distance between the ring blob and the
black hole blob, which is given by \rradangmom. Since $j_L - d^I \widehat Q_I$
is quantized, the minimal size of the ensemble of GH points is given by:
\eqn\rminbase{r|_{\rm min} \approx {1 \over d^1+d^2+d^3} \,.}
More generically, in the scaling limit, the GH size of a solution with left-moving angular momentum $J_L$ is
\eqn\rminbas{r|_{\rm min} \approx {J_L \over d^1+d^2+d^3} \,.}

Since the $d^I$ scale like the square-roots of the ring charges, we
can see that in the classical limit, $r|_{\rm min}$ becomes zero and the
throat becomes infinite.

\subsec{Numerical results for a simple merger}

Given that most of the numerical investigations and most of the
derivations we have discussed above use black hole microstate made
from a very large number of points, it is quite hard to illustrate
explicitly the details of a microstate merger.

To do this, we investigate a black hole microstate that is made from three points, of
GH charges $-n$, $2n+1$, and $-n$, and its merger
with a black ring microstate of GH charges $-Q$ and $+Q$. This black hole microstate can be obtained by redistributing the position of the
GH points inside the BH blob considered in the previous subsection,
putting all the $+1$ charges together
and putting half of  the $-1$ charges together on one side of the positive center
and the other half on the other side\foot{Since the $k$ parameters on the BH
points are the same, the bubble equations give no obstruction to moving BH
centers of the same GH charge on top of each other.}

We consider a configuration with five GH centers of charges
\eqn\ghchn{q_1=-12\,, \quad q_2=25\,, \quad q_3=-12\,, \quad q_4=-20\,,
\quad q_5=20\,.}
The first three points give the black hole ``blob,'' which can be thought as coming
from a blob of $N-2 = 49$ points upon redistributing the GH points as described
above.  The $k^I$ parameters of the black hole points are:
\eqn\ghch{k_1^I = |\,q_1| \, \hat k_0^I~,~~~k_2^I = |\,q_2| \, \hat k_0^I~,~~~k_3^I
= |\,q_3| \, \hat k_0^I~,}
%
%
%
where $ \hat k_0^I $ is the average of the $k^I$ over the BH points,
defined in \kIavg. To merge the ring and the black hole microstates we
varied $ \hat k_0^2 $ while keeping $ \hat k_0^1 $ and $ \hat k_0^3 $ fixed:
\eqn\khatnum{ \hat k_0^1 ={5 \over 2}\,, \qquad  \hat k_0^3 ={1\over 3},~  }
We also fixed the ring parameters $f^I $ and $d^I$ as follows:
\eqn\fdnum{d^1=100\,, \quad d^2=130\,, \quad d^3=80\,, \qquad f^1=f^2=160\,,
\quad f^3=350  }
The relation between these parameters and the $k^I$ of the ring is given
in \ringchgs, where $N-2$ (the sum of $|q_i|$ for the black hole points) is now
$|q_1| +|q_2|+ |q_3| = 49$.


\def\nicespacea#1{{~~#1~~}}
\goodbreak
{\vbox{\ninepoint{
$$
\vbox{\offinterlineskip\tabskip=0pt
\halign{\strut\vrule#
&\vrule\hfil #\hfil
&\vrule\hfil #\hfil
&\vrule\hfil #\hfil
&\vrule\hfil #\hfil
&\vrule\hfil #\hfil
&\vrule\hfil #\hfil
&\vrule\hfil #\hfil
&\vrule\hfil #\hfil
&\vrule\hfil #\hfil
\cr
\noalign{\hrule}
&~ Parameters~ &\nicespacea{$ \hat k^2_0$}&\nicespacea{$x_4-x_3$}&
\nicespacea{$\displaystyle{x_4-x_3 \over x_2-x_1}$}&
\nicespacea{$\displaystyle{x_2-x_1\over  x_3-x_{2}}$}
&\nicespacea{$\displaystyle{x_2-x_1\over x_5-x_4}$}&\nicespacea{$J_L$}&
\nicespacea{$\displaystyle{Q_1 Q_2 Q_3-{J_R^{2^{{\White 9^1 \Black}}}}\!\!\!\! /4 \over Q_1 Q_2 Q_{3_{\White 9_9 \Black}}}$}&
\cr
\noalign{\hrule height1pt}
&~1~& 3.0833 &~175.5~ & 2225 & 1.001 & 2.987 &~215983~&.275
&
\cr
\noalign{\hrule }
&~2~& 3.1667 &23.8 & 2069 & 1.001 & 3.215 &29316&.278
&
\cr
\noalign{\hrule }
&~3~& 3.175 &8.65 & 2054 & 1.001 & 3.239 &10650&.279
&
\cr
\noalign{\hrule }
&~4~& 3.1775 &4.10 & 2049 & 1.001 & 3.246 &5050&.279
&
\cr
\noalign{\hrule }
&~5~& 3.178 &3.19 & 2048 & 1.001 & 3.248 &3930&.279
&
\cr
\noalign{\hrule }
&~6~& 3.17833 &2.59 & 2048 & 1.001 & 3.249 &3183&.279
&
\cr
\noalign{\hrule }
&~7~& 3.17867 &1.98 & 2047 & 1.001 & 3.250 &2437&.279
&
\cr
\noalign{\hrule }
&~8~& 3.1795 &.463 & 2046 & 1.001 & 3.252 &570&.279
&
\cr
\noalign{\hrule }
&~9~& ~3.17967~ &.160 & 2045 & 1.001 & 3.253 &197&.279
&
\cr
\noalign{\hrule}}
\hrule}$$
\vskip-7pt
\leftskip 2pc
\rightskip 2pc\noindent{\ninepoint\sl \baselineskip=8pt
{\bf Table 1}: Distances between points in the scaling regime. The
value of $\hat k_0^2$ is varied to produce the merger, and the other
parameters of the configuration are kept fixed: $ Q=20,~q_1=q_3=-12,~q_2=25,~\hat
k_0^1={5 \over 2},~\hat k_0^3={1\over
3},~d^1=100,~d^2=130,~d^3=80,~f^1=f^2=160,~f^3=350$. Both the charges
and $J_R$ remain approximately the same, with $J_R \approx 3.53 \times
10^7$.
}
}
\vskip7pt}}

The charges and $J_R $ angular momentum of the solutions are approximately
\eqn\qjnum{Q_1 \approx 68.4 \times 10^3,~Q_2 \approx 55.8 \times 10^3,~
Q_3 \approx 112.8 \times 10^3,~J_R \approx 3.53 \times 10^7,}
while $J_L$ goes to zero as the solution becomes deeper and deeper.  The
result of the merger is a microstate of a BMPV black hole that is $28\%$ below
maximal rotation (see the last column of Table 1.)

Numerically solving the bubble equations \bubbleeqns, one obtains the
positions $x_i$ of the five points as a function of $\hat k^2_0$. Some of
the results are shown in Table 1.  As one can see from the table, a very small
increase in the value of
$\hat k^2_0$ causes a huge change in the positions of the points on
the base. If we were merging classical black holes and classical black rings,
this increase would correspond to the black hole and the black ring
merging. For microstates, this results in the scaling described
above: all the distances on the base become smaller, but their ratios
remain fixed.

Analytically verifying that these solutions have no closed timelike
curves is not that straightforward, since the quantities in \noCTCs\
have several hundred terms. However, we have investigated numerically
and graphically  for the possible presence of closed timelike curves, and have found
that the equations \noCTCs\ are satisfied throughout the scaling solutions
discussed in this sub-section.

\subsec{A pincer movement: Two rings and a black hole}

The scaling solutions that we have constructed so far do not
have $J_L =0$, but only achieve this in the extreme limit in which
$r_0 \to 0$.  Thus these solutions have the charges of a  black hole of
non-zero entropy only in the large-throat limit.  While this feature of deep microstates
makes them  similar  to typical microstates, it is interesting to explore whether solutions
with $J_L = 0$  and $Q_1 Q_2 Q_3 > {J_R^2/4}$ must necessarily have long throats,
or whether they can also  be shallow.

Constructing solutions with $J_L =0$ is very simple:  One can put identical ring blob on opposite sides of the black-hole blob.  Indeed, one
can make $\ZZ_2$ invariant solutions in this manner and, as we remarked
earlier, this guarantees $J_L =0$.  From the perspective of the asymptotically $\IR^4$ base,  a ring blob located on axis to the left of the black-hole blob
describes a lack ring in one $\IR^2$ plane in $\IR^4$, while a ring blob located
on axis to the right of the black-hole blob describes a ring in the
orthogonal $\IR^2$ plane.  We are thus adding two identical, perpendicular
rings  to the black hole, thereby guaranteeing $J_1 =J_2$.

We will again consider the merger of these rings with the black hole,
and show that having two rings instead of one modifies the analysis
above very little.  We can again find a scaling behavior, but now $J_L
= 0$ and $Q_1 Q_2 Q_3 > {J_R^2/4}$ {\it throughout the merger}. Hence, this
microstate will have the charges of a black hole of non-zero entropy when
it is shallow  (before the scaling regime), when it is deep, and in the
intermediate regime.

Once again we consider a system of $N$ GH points but now the
first pair and last pair are ``exceptional,'' and correspond to rings.
The black-hole blob consists of the $(N-4)$ points in the middle:
\eqn\pincerq{q_1 = Q_1\,, \quad  q_2 = - Q_1 \,, \quad q_{N-1} = - Q_2\,,
\quad  q_N = Q_2 \,,   \quad q_j = (-1)^{j+1}\,, \quad  j=3,\dots,N-2 \,,}
\eqn\pincerkbh{k_i^I ~=~  k^I\,, \qquad j~=~3, \dots, N-2 \,.}
As before, it is convenient to define\foot{These definitions are not
naturally orthogonal fluxes in cohomology, but they have the
virtue of being symmetric between the rings.}:
\eqn\ddefns{\eqalign{d_1^I ~\equiv~& 2 \, \big(  k_{1}^I + k_{2}^I \big) \,, \qquad
f_1^I ~\equiv~ 2 \,(N-4) \,   k^I +  \big(1 + \coeff{1}{Q_1}\big)\, k_2^I  +
 \big(1 - \coeff{1}{Q_1}\big)\, k_1^I \cr
 d_2^I ~\equiv~&  2 \, \big(  k_{N-1}^I + k_{N}^I \big) \,, \qquad
f_2^I ~\equiv~ 2 \,(N-4) \,  k^I +  \big(1 + \coeff{1}{Q_2}\big)\, k_{N-1}^I  +
 \big(1 - \coeff{1}{Q_2}\big)\, k_{N}^I  \,.}}
We now have
\eqn\QJhat{\widehat Q_I ~=~ 2\,(N-3)(N-5)\, C_{IJK}\, k^J k^K \,, \qquad
\widehat J_R~=~ \coeff{8}{3}\,(N-3)(N-4)(N-5)\, C_{IJK}\, k^I k^J k^K  \,,}
and we define
\eqn\Qringi{Q^{Ring,i}_I   ~\equiv~   \, C_{IJK}\,  d_i^J f_i^K   \,.}
\eqn\JringRi{j_{R,i}  ~\equiv~ \coeff{1}{2} \, C_{IJK}\, \big(f_i^I f_i^J d_i^K ~+~
f_i^I  d_i^J  d_i^K\big)  ~-~ \coeff{1}{24} \,(1-Q_i^{-2})\, C_{IJK}\,  d_i^I d_i^J d_i^K \,.}
Then we find:
\eqn\Qtot{Q_i ~=~ \widehat Q_I ~+~  Q^{Ring,1}_I  ~+~ Q^{Ring,2}_I  ~+~
C_{IJK}\,  d_1^J d_2^K   \,.}
and
\eqn\JRtot{\eqalign{J_R  ~=~ \widehat J_R  ~+~    j_{R,1} ~+~ j_{R,2} &~+~
(d_1^I + d_2^I)\, \widehat Q_I  ~+~  d_1^I \,Q^{Ring,2}_I   ~+~
 d_2^I \,Q^{Ring,1}_I  \cr & ~+~  \coeff{1}{2} \, C_{IJK}\,
 (d_1^I d_2^J d_2^K + d_2^I d_1^J d_1^K  )  \,.}}
These formulas agree identically to the formulas that give the charges and angular
momenta of two concentric black rings with a black hole in the middle \GauntlettQY.
One can also obtain similar expressions for $J_L$, but here we  will focus
on  $\ZZ_2$-invariant solutions and so $J_L=0$.

Now set
\eqn\symmfluxes{d_1^I ~=~ d_2^I ~=~ d ^I \,, \qquad f_1^I ~=~ f_2^I ~=~ f ^I\,,}
and impose $\ZZ_2$ symmetry in the distribution of GH points. In particular,
take the origin to be the center of the distribution and define:
\eqn\rDeltadefn{ r_0 ~\equiv~ |\vec r_{N-1}|  ~=~ |\vec r_{2}| \,, \qquad
\Delta ~\equiv~  |\vec r_{N } - \vec r_{N-1}|~=~  |\vec r_{1 } - \vec r_{2}| \,.}

The analysis for $\Delta, r_i \ll r_0$ proceeds in an almost identical manner
to the asymmetric solution described above, but with three modifications:
\item{(i)} One shifts $ N \to N-2$ throughout because one now has $N-4$
points, as opposed to $N-2$ points, in the black-hole blob.
\item{(ii)}  The value of $r_0$ is still given by \newringrad\
and the merger condition is still given by \exactMcond, but this is
no longer related to $J_L$ since the latter is always zero.
\item{(iii)}  While the expression for $r_0$ is unchanged, the black-hole
blob now feels twice the compression because there are now two rings.
Thus \unieqns\ becomes:
\eqn\newunieqns{\mu_i  ~\approx~
\Big(\sum_{I=1}^3 \, k^I \Big)  ~+~ {2  \over  r_0} \,C_{IJK} \,
\Big[ \coeff{1}{8}\,  d^I\, k^J\, X^K  ~-~ \alpha_0^{-1}\,Q\, \gamma\, \big(
\coeff{1}{32}\, X^I  \,  X^J  \,k^K ~+~ \coeff{1}{6}\, k^I\, k^J\,k^K \big) \,   \Big]   \,.}

Thus to leading order, the two rings do not influence each other and
each settles down as if the other were not there.  However the black hole
shrinks to about half its previous size.  From numerical simulations,
we see some amusing features at  sub-leading order.  In particular, when a
single ring is present, the distribution of GH points in the black-hole blob
develops a small dipolar asymmetry and the scale factor varies very slightly across
the blob (to first sub-leading order).  When two rings are present, the
symmetry is restored and the variation in the distribution of GH points
only appears in the quadrupole moments and the scale factor has only
tiny variations  across the blob ({\it i.e.} to second sub-leading order).

\subsec{The metric structure of the deep microstates}

The physical metric is given by \fullmet\ and the physical lengths on
the three-dimensional base of the GH space are therefore determined by:
\eqn\physlen{ds_3^3 ~=~ (Z_1 Z_2 Z_3)^{1/3} \, V \, d \vec y \cdot d \vec y \,.}
The physical lengths are thus determined by the functions,
$Z_I V$, and if one has:
\eqn\AdSthroat{(Z_1 Z_2 Z_3)^{1/3} \, V  ~\sim~ {1 \over r^2} \,,}
then the solution looks is an $AdS_2 \times S^3$ black hole throat. In the region
where the constants in the harmonic functions become important, this throat turns
into an asymptotically flat  $\IR^{(4,1)}$ region.  Near the GH centers that give the
black hole bubbles the function $Z_1 Z_2 Z_3$ becomes constant.  This corresponds
to the  black-hole throat ``capping off''.   As the GH points get closer in the base, the region where
\AdSthroat\ is valid becomes larger, and hence the throat becomes longer.

As one may intuitively expect, in a scaling solution the ring is always in the
throat of the black hole. Indeed, the term ``1'' on the right hand side of \scalefac\
originates from the constant terms in $L_I$ and $M$, defined in \LMform. In the
scaling regime this term is subleading, which implies the ring is in a region
where the $1$ in the $L_I$ (and hence the $Z_I$) is also subleading.  Hence, the
ring lies in the $AdS$ throat of the black-hole blob.

Increasing the scale factor, $\lambda$, in \scalefac\ means that
the bubbles localize in a smaller and smaller region of the GH base,
which means that the throat is getting longer
and longer.  The physical circumference of the throat is fixed by the
charges and the angular momentum, and remains finite even though the blob
is shrinking on the GH base. Throughout the scaling the throat becomes deeper and deeper; the ring remains in the throat, and also
descends deeper and deeper into it, in direct proportion to the overall
depth of the throat.

On a more  mechanistic level, the physical distance through the blob
and the physical distance from the blob to the ring are
controlled by  integrals of the form:
\eqn\physsize{ \int \, (Z_1 Z_2 Z_3 \, V^3 )^{1/6}  \, d \ell \,.}
In the throat the behavior of this function is given by \AdSthroat\
and this integral is logarithmically divergent as $r \to 0$.  However,
the $Z_I$ limit to finite values at $\vec r = \vec r_j$  and between two very
close, neighboring GH points in the blob, the integral has a dominant
contribution  of the form
\eqn\lowerbd{ C_0 \ \int \,|(x-x_i)(x-x_{j})|^{-1/2}   \, d x \,,}
for some constant, $C_0$, determined by the flux parameters.  This
integral is finite and indeed is equal to $C_0 \, \pi$.   Thus we see
that the throat gets very long but then caps off with bubbles of finite
physical size.

Since the ring carries charge, its presence will cause
the limiting value of the throat size to jump.  Figure 4 shows a
representative numerical example.  These are plots showing $\log(Z_I V)$
against $\log(r)$.  In the Coulomb region, away from the ring and the
blob, one has
\eqn\ZVasymp{Z_I \, V ~\sim~ {Q_I \over r^2} \,,}
where $Q_I$ is the charge of the configuration inside a radius $r$.
Therefore the log-log plot will show such regions as straight lines
of gradient $-2$ with an intercept fixed by $\log(Q_I)$.  Both graphs in
Figure 4 show a transition between two such lines, and the ``jump'' between
the lines is accounted for by the jump in the charge as one
encounters the ring.   The first graph shows the behavior of $Z_I V$
taken right through the GH points of the ring, while the second
graph shows the behavior $Z_I V$ in an orthogonal direction in which one
does not pass near the GH points of the ring.  As expected, in both instances,
the throat widens as one passes the ring radius.

\goodbreak\midinsert
\vskip .2cm
\centerline{ {\epsfxsize 2.4in\epsfbox{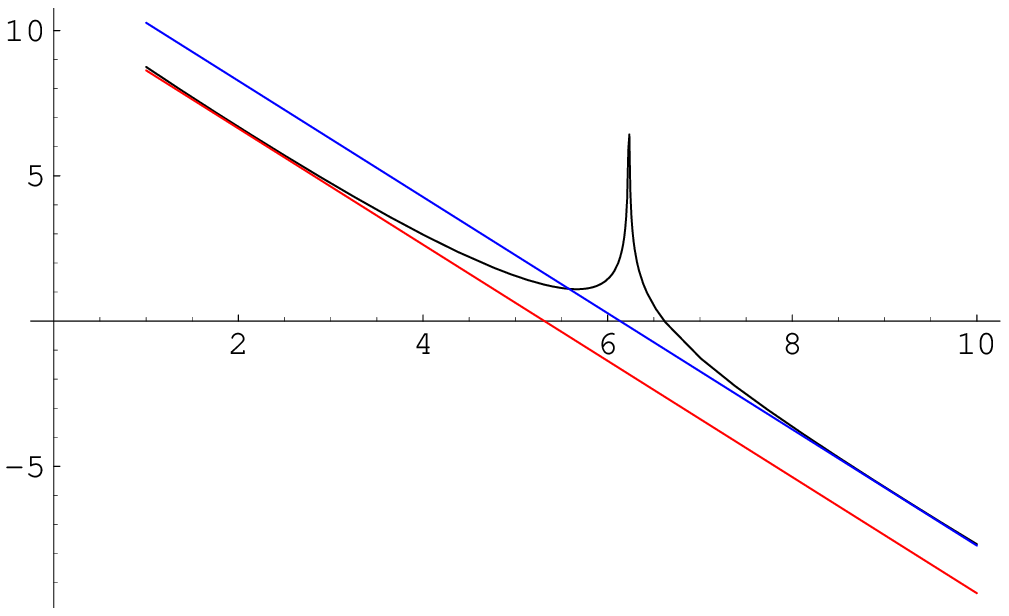}} \hskip 0.5in
{\epsfxsize 2.4in\epsfbox{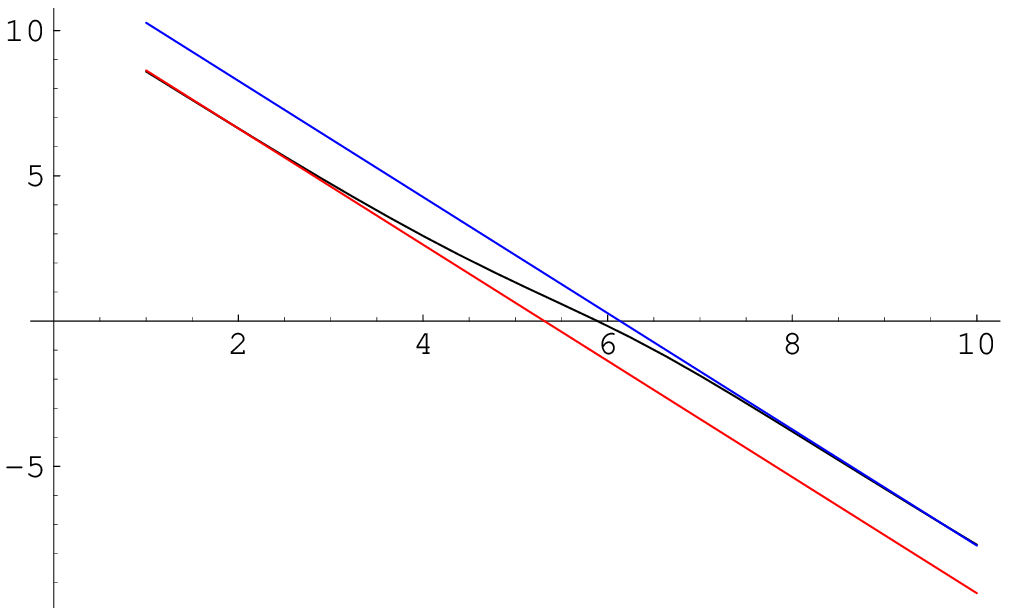}}}
\vskip -.25cm
\leftskip 2pc
\rightskip 2pc\noindent{\ninepoint\sl \baselineskip=8pt
{\bf Fig.~4}:  Typical plots of $\log(Z_I V)$ against $\log(r)$. The straight
lines depict exact Coulomb behavior for $Z_I$ while the actual solutions transition
between two such lines as the ring charge is included.  The first graph
passes directly through the ring points, while the second goes in an
orthogonal radial direction.}
\endinsert

\newsec{Reversible Mergers}

\subsec{Some numerical results in the merger transition}

Once again we performed many numerical simulations for
an odd number,  $N$,  of co-linear  GH charges.
As before, we focused on a system with GH charges $q_j = (-1)^{j+1}$, $j=1,\dots, N-2$ and
$q_{N-1} = -Q$, $q_{N} = Q$.  We only considered solutions in which
the charges  alternate in sign as one goes along the distribution.
Before we discuss the results we first need a better
geometric understanding of this configuration.

We will take $(y_1,y_2, y_3) = (x,y,z)$ and suppose that
the line of charges lies along the $z$-axis, with the
$j^{\rm th}$ point at $z = z_j$.    We consider solutions
with $z_{j+1} > z_j$ so that the charges alternate in sign.
We will also denote the  polar angle in the  $(x,y)$ plane by $\phi$.
One should recall that the GH metric requires one to solve for
$\vec A$ in \Vform\ and for the configuration we are considering
we can take
\eqn\Aform{ \vec A  \cdot d \vec y~=~ \sum_{J=1}^N \,  q_j
{(z -z_j) \over r_j}  \, d \phi \,.}
On the $z$-axis, $r_j = |z - z_j| $, and since the charges alternate
one has, on the $z$-axis:
\eqn\Aaxis{ \eqalign{\vec A  \cdot d \vec y~=~&  (-1)^\ell \, d \phi \,, \quad
{\rm for }  \quad  z_{\ell-1} < z  <z_{\ell} \,, \  \  \ell \ne N \cr
\vec A  \cdot d \vec y~=~&  -(2Q-1)\, d \phi \,, \quad
{\rm for }  \quad  z_{N -1} < z  <z_{N}   \,,}}
where, for the purposes of the inequality, $z_0 = -\infty$, $z_{N+1} = +\infty$.

Now recall that the change of variables at infinity that takes the GH metric
with $V \sim {1 \over r}$ to flat $\IR^4 = \IR^2 \times \IR^2$ with coordinates
$(u, \theta_1)$ and $(v, \theta_2)$ in each of the $\IR^2$ planes involves
taking:
\eqn\thetavars{\theta_1 ~=~ \coeff{1}{2} \, (\psi +\phi) \,, \qquad
\theta_2 ~=~ \coeff{1}{2} \, (\psi -\phi) \,.}
This, combined with \Aaxis, means that the GH fiber defined by $(d \psi +A)$
alternates between being in the $(u, \theta_1)$ plane and the $(v,\theta_2)$
plane.  In particular, if the two exceptional points are far away, then there
is a long interval ($z_{N-2} < z < z_{N-1}$) with GH fiber in the $(u, \theta_1)$ plane.
Once one pushes this through the complete change of variables,
one finds that, at large scales,  such a ring blob resembles a supertube in the
$(u, \theta_1)$ plane.
More generally,   whether a blob like this corresponds to a ring in
the $(u, \theta_1)$ plane or the $(v, \theta_2)$ plane depends upon
whether the large interval on the $z$-axis has $(d \psi +A)$
equal to $(d \psi + d \phi)$ or $(d \psi  - d \phi)$, respectively.

In the simulations we set all the $k^I_j = +1$ in the initial blob
($j =1, \dots, N-2$), and took $d^I = d$ and $f^I =f$, but otherwise
arbitrary.   We took $Q$ to be fairly large (usually about $30$).
We then did many series of simulations in which we adjusted
$d$ but kept the ratio, $f/d$, fixed.  The ring width,
$\Delta$, becomes smaller when the values of $f/d$ get larger.
The value of $d$ was adjusted  so as to move the two exceptional
``black ring'' points in from a great distance from the black-hole blob.   The numerical
solutions tend to be very unstable, or rather delicate, particularly when the
black ring points are near the bubbling black hole.  This is because
there are a great many solutions to the bubble equations in which points
are re-ordered.  These solutions are very ``close to one another'' when
bubbles become small, and the numerical algorithms readily jump between
different branches of the solution space.  In spite of this, if one carefully adjusts
the solution  adiabatically and uses the previous  solution as initial data
to find the next solution, one can follow a single branch in the solution
space.

At merger (defined by $J_1 = J_2$) we found that the assumption that the two ring
points are at a distance from the blob that is of the same order as the size of $r_0$ in \rradangmom\ is no longer valid.
It turns out that as one comes close to
satisfying the merger condition, $r_0$ is still somewhat larger than
$r_i$, but $\Delta/r_0$ is extremely small.   Dropping the $\Delta/r_0$
term in the derivation of the  estimate of $r_0$ means that when the
merger condition ($J_1 = J_2$) is satisfied, one still has:
\eqn\mergeringrad{r_0 ~\approx~  Q^{-1} \,  \bigg[24\, \sum_I \, d^I\bigg]^{-1}
 \, C_{IJK} \, d^I d^J d^K  \,.}
While we have not been very careful in the derivation of this estimate, it
turns out to work rather well in the numerical simulations of
mergers with parallel fluxes. For example, for the result depicted in the first graph
of Figure 5, we found that the exceptional points were $37.95$ units
(taking $k=1$) from the center of the blob.  The formula, \mergeringrad,
gives $r_0 \approx 36.68$.   We also found that at the merger value
one could bring the exceptional points arbitrarily close to the edge of the blob
by taking $Q$ to be large enough.   This is also consistent with \mergeringrad,

In all the numerical simulations, the blob maintained the BMPV profile
of Figure 3  so long as the exceptional points remained relatively far away.
However, as the exceptional points approach the blob closely, the blob
undergoes a ``phase transition.''  If the merger value of $r_0$ is still
significantly larger than the blob size (as it is for the results depicted
in Figure 5) then bringing the exceptional points near the blob entails
tracking the solution past the merger condition.  That is, one follows
the solution as $J_1 -J_2$ passes from positive to negative values.
We found that the two exceptional  points generically do not enter
the blob (at least on this branch of solution space) but get very close to it, and
actually ``steal'' the outermost GH center and form a cluster of three points of net
GH charge $+1$.  This cluster then moves away back towards the
original distant position of the exceptional points.  Throughout this
process $J_1 - J_2$ monotonically decreases.

The ``theft'' of a $+1$ GH point renders the blob neutral, and makes it to change
from a  black-hole blob to a  ring blob,  with
the concomitant redistribution of dipole pairings.  This ``unzipping''
phase transition in the blob happens extremely quickly as one
varies $d$ below the merger point.  Typical results
are depicted in Figure 5.

\goodbreak\midinsert
\vskip .2cm
\centerline{ {\epsfxsize 2.4in\epsfbox{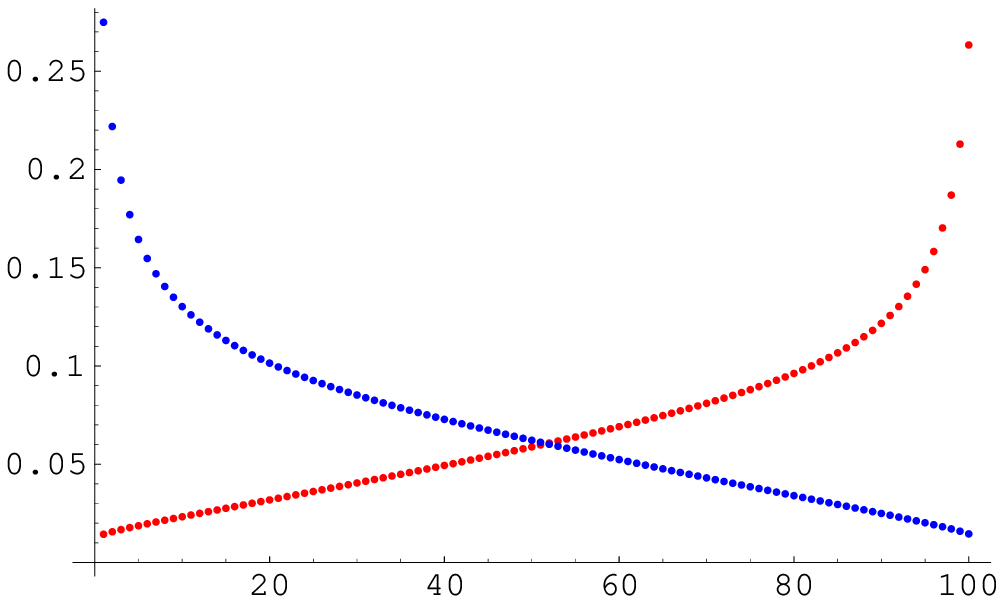}} \hskip 0.5in
{\epsfxsize 2.4in\epsfbox{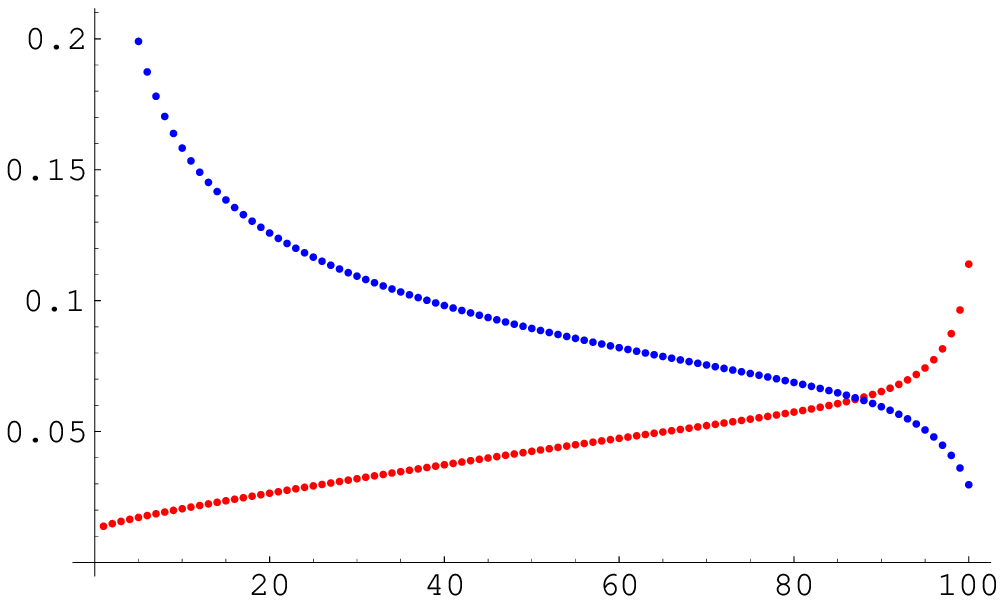}}}
\vskip 0.2in
\centerline{ {\epsfxsize 2.4in\epsfbox{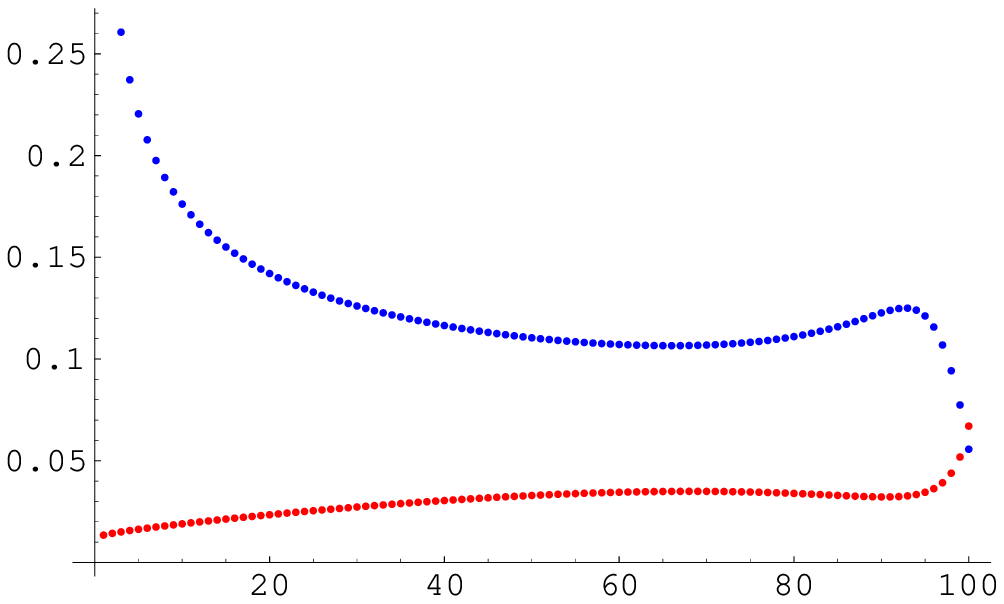}} \hskip 0.5in
{\epsfxsize 2.4in\epsfbox{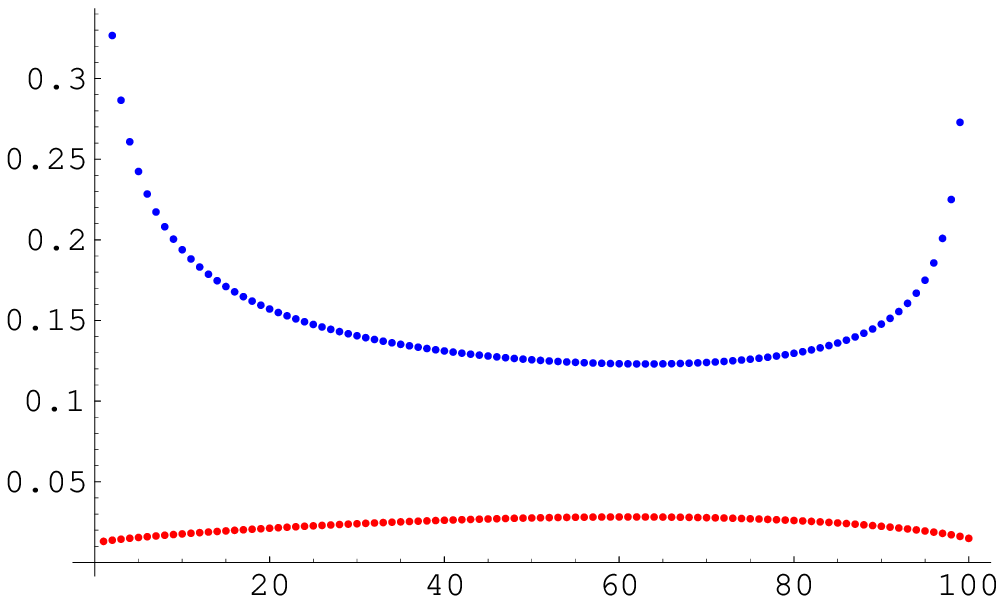}}}
\vskip -.25cm
\leftskip 2pc
\rightskip 2pc\noindent{\ninepoint\sl \baselineskip=8pt
{\bf Fig.~5}: These graphs show the distances between neighboring
GH points  for $N = 203$, $Q =30$, $ f= 4d$, but with $d$ decreasing from its
merger value, $d = 114.9074$, in the first graph to $d = 114.84$ in the top-right,
$d = 114.825$ in the bottom-left and $d = 114.80$ in the bottom right. Note
how the distributions change between that of a  black-hole blob  to that
of a  ring blob as $d$
changes only by $0.1\%$. In the first graph the blob diameter is $14.02$ while
the space to the first exceptional point is $|z_{N-2} - z_{N-1}| = 30.94$. In the
second graph these quantities are $14.46$ and $0.999$, in the third they are
$16.02$ and $0.547$. In the  fourth graph, the ``theft'' has happened:  The blob of
$N-3$ points has radius $17.55$ and the ``large
distance'' is now $|z_{N-3} - z_{N-2}| = 4.908$.}
\endinsert

Before the transition the ``large interval'' lies between $z_{N-2}$
and $z_{N-1}$, and so the black ring points correspond to a ring in the
$(u, \theta_1)$ plane. After the transition, as the cluster of three
charges moves away, the ``large interval'' now lies  between $z_{N-3}$
and $z_{N-2}$. Therefore, at large scales the resulting configuration corresponds
to a black hole microstate surrounded by a black ring microstate in
the $(v, \theta_2)$ plane!   Thus we are exploring a branch of the
solution space in which a bubbling supertube in one plane collapses
into an extremal black hole and a then a supertube emerges from the
extremal black hole, but now in the orthogonal plane.  All the while,
$J_1 -J_2$ decreases.  These obviously all represent reversible mergers.

There are evidently a vast number of solutions and probably a vast number
of branches in the moduli space of locations of GH points.  The branch we explored was
probably the simplest that allowed $J_1 -J_2$ to decrease smoothly
while preserving the order and co-linearity of the GH points.
There are almost certainly many other solutions in which the GH points
are reordered, perhaps with the exceptional points penetrating
more deeply into the blob.  There are also quite probably solutions
in which more GH points are stolen from the blob.

\subsec{Estimating the size of shallow microstates from mergers}

As we have seen is section 5, deep microstates have very long
black-hole-like throats, and it is clear that they have the same size
and macroscopic features as the black hole. However, it is interesting
to explore whether ``shallow'' microstates, that do not have long
throats, also have the same macroscopic properties as the actual black
hole. A  good way to estimate this size is to probe the shallow microstates using mergers.

The irreversible mergers that we considered in section 5 can, by no means, be considered ``probing''.   The merger collapses the foam, resulting in a deep
microstate.  This is partially because the ``probing ring'' had comparable
charges to those of the black hole.
%
%
 To find the ``size'' of a shallow microstate one
needs to merge it with a small probe-like ring microstate and perform as
nearly a reversible merger as possible.  The calculations of the previous
sub-section describe exactly such a process.

As we have seen, the reversible merger happens when the ring
foam sits in the vicinity of the black hole foam. Therefore, at the
merger point the shallow microstates essentially touch, resulting in
another shallow microstate.  In the merger of classical black rings
and classical black holes, the horizons of the two objects also touch
at the merger point. Hence, reaching the horizon of the black hole
corresponds in the microstate picture to reaching the ``edge'' of the
shallow microstates. This indicates that shallow microstate geometries also have the same
size and macroscopical properties as their corresponding classical
black holes.

\newsec{Deep Microstates and Typical Microstates}

As we have seen in the previous section, the throats of the deep microstates
become infinite in the classical limit. Nevertheless, taking into account
flux quantization one can find that the GH radius of microstates does not go
all the way to zero, but to a finite value \rminbase, which corresponds to setting $J_L=1$.

One can estimate the energy gap of the solution by considering the
lightest possible state at the bottom of the throat, and estimating
its energy as seen from infinity.  The lightest massive particle one
can put on the bottom of the throat is not a Planck-mass object, but a
Kaluza-Klein mode on the $S^3$. Its mass is
\eqn\massKK{m_{KK} = {1 \over R_{S^3}} = {1 \over (Q_1 Q_2 Q_3)^{1\over 6} }}
and therefore the mass gap in a microstate of size $r_{\rm min}$ in the GH base is:
\eqn\gapmicro{\Delta E_{r_0} = m_{KK} \sqrt{g_{00}}|_{r=r_{\rm min}}
=  m_{KK} (Z_1 Z_2 Z_3)^{-1/3}|_{r=r_{\rm min}} = {r_{\rm min} \over (Q_1 Q_5 Q_P)^{1/2}}.}
Since in the case of a ring-hole merger $r_{\rm min}$ depends on the
sum of the $d^I$, its relation with the total charges of the system is
not straightforward. Nevertheless, we can consider a regime where
$Q_1\sim Q_5 > Q_P$, and in this regime the dipole charge that
dominates the sum in \rminbas\ is $d^3 \approx \sqrt{ Q_1 Q_5 \over Q_P}$. Hence
\eqn\rminlim{r_{\rm min} = {J_L \over d^3} \approx J_L \sqrt{Q_P \over Q_1 Q_5}~, }
giving
\eqn\gap{\Delta E_{r_0} \approx {J_L \over Q_1 Q_5}~.}
For $J_L=1$ this matches the charge dependence of the mass gap of the black hole \MaldacenaDS.

This M-theory frame calculation is done in the limit $Q_1 \sim Q_5 > Q_P$, which is the limit in which the solution, when put into the
D1-D5-P duality frame, becomes asymptotically $AdS_3 \times S^3 \times T^4$. \foot{As shown in \BenaTK, in this limit $d^1+d^2+d^3 \rightarrow d^3$, which justifies going from \rminbas\ to \rminlim.} Hence, the mass gap computed in the bulk \gap, should match the
mass gap of the dual microstate in the D1-D5 CFT.

As it is well known (see \refs{\AharonyTI,\DavidWN } for reviews)
the states of this CFT can be
characterized by various ways of breaking an effective string of
length $N_1 N_5$ into component strings.  BPS momentum modes on these
component strings carry $J_R$. The fermion zero modes of each
component allow it in addition to carry one unit of $J_L$.
The typical CFT microstates that contribute to the entropy of the
three-charge black hole have one component string
\refs{\BreckenridgeIS}; microstates dual to objects that have a
macroscopically large $J_L$ have the effective string broken into many
component strings \refs{\LuninJY,\LuninIZ,\BenaTK }. Hence, the only
way a system can have a large amount $J_L$ is to be have many
component strings.  The CFT mass gap corresponds to exciting the longest
component string, and is proportional to the inverse of its length.

The formula \gap\ immediately suggests a dual for the deep microstates.
Consider a long effective string of length $N_1 N_5$ broken
into $J_L$ component strings of equal length.  Each component string can carry
one unit of left-moving angular momentum, totaling up to $J_L$. The length
of each component string is
\eqn\lbit{l_{\rm component} = {N_1 N_5 \over J_L}~,}
and hence the CFT mass gap is
\eqn\cftgap{\Delta E_{CFT} \approx {J_L \over N_1 N_5}~.}
This agrees with {\it both} the $J_L$ dependence and the charge dependence of the bulk mass gap.
While we have been cavalier about various numerical factors of order one,
the agreement that we have found suggests that deep microstates of angular
momentum $J_L$  are dual to CFT states with $J_L$ component strings.
If this is true, then the deepest microstates, which have $J_L = 1$, correspond
to states that have only one component string, of length $N_1 N_5$. This is a
feature that typical microstates of the three-charge black hole have, and the
fact that deep microstates share this feature is quite remarkable.

Our analysis here has been rather heuristic.  It would be very interesting to
examine this issue in greater depth by finding, at least approximate solutions
to the wave equation in these backgrounds, or performing a time-of-flight analysis
along the lines of  \refs{\LuninJY,\LuninIZ,\GiustoIP}.

\newsec{Final Remarks}

We have examined bubbled solutions constructed using a generalized
Gibbons-Hawking base space.  We have shown that if the GH points are
localized in blobs of total GH charge $+1$, then the solutions
correspond to microstates of the maximally-spinning BPS black hole.
Similarly, blobs whose total GH charge is zero correspond to
zero-entropy black ring microstates.

We found numerically that inside a bubbling black ring the GH centers
form tight neutral dipole pairs, separated from each other by
relatively large distances.  On the other hand, inside the black hole, the dipoles
are only tightly bound around the edges, and in the middle one finds
an equally spaced distribution of positive and negative GH charges.

We have then investigated the merger of bubbling black rings and
bubbling black holes, and have found two types of mergers. If the
final state has the charges corresponding to a black hole of
classically large horizon area (this would correspond to an
irreversible merger of a black hole and a black ring), then during the
merger the distances between the GH points on the base shrink, while
the ratios of all these distances remain fixed. The
physical size of the bubbles that form the solution remains the same, but
the throat of the solutions becomes deeper and deeper. We have therefore
called these solutions ``deep'' microstates.

Classically this scaling goes on forever, until all the distances in the base are zero.
However, the fact that the dipole charges are quantized does not allow
one to take them exactly to the value which would correspond to an
infinite throat; hence, the depth of the microstate is very large but finite for
finite charges, and only becomes infinite in the classical limit.

When the length of the throat of a deep microstate is very large, this solution
more and more accurately resembles a genuine BPS back hole (which has
an infinite throat), except for the  presence of a horizon. This feature of the
deep microstates makes them very attractive candidates for being typical BH microstates.

The similarity of the deep microstates to the typical CFT microstates can also be
observed from the crude $AdS$-CFT analysis
we have attempted in Section 7.  We have argued that deep microstates have the right mass gap to be dual to  CFT states described by one long component string, and are thus
similar in this respect to the typical microstates. It is quite remarkable that the calculation that relates deep microstates of angular momentum $J_L$ to states with $J_L$ component strings is independent of the number of GH centers.  One can speculate that solutions with different numbers of GH points may be related to CFT states that have different distributions of momentum modes on the component strings. It would be very important to establish whether such a relation exists, and more generally to undertake an in-depth CFT investigation of deep microstates.  This will be very fruitful, and
will hold the key to relating bubbling microstates to their CFT counterparts.

We have also used the merger of a black hole with {\it two} concentric black rings
to obtain microstate solutions that correspond to a black hole with classically large entropy
but which can range between shallow and deep microstates.  Since the redshift of
the shallow and intermediate solutions does not diverge in the classical limit, it is likely that they are
``less typical''  than the deep microstates, or that they are not microstates of the black hole, but microstates of a configuration
of a black hole with two concentric black rings.
 It would be very interesting to explore how
generic these solutions are, and what their CFT dual is. It would also be very interesting
to try to construct deep black ring microstates (which would be similar to typical
microstates of a black ring), and to analyze the merger of two deep microstates,
or of a shallow and a deep microstate.

We have also analyzed mergers in which no scaling takes place, and which
result in ``shallow'' microstates. We have found that during these mergers two
shallow microstates join, forming a larger shallow microstate. We have used
this fact to estimate the size of shallow microstates, and to argue that for the maximally-spinning black hole, the bubbles extend to the location of
 the would-be classical horizon.  This is a
 strong indication that the shallow microstate geometries also have
features that typical BH microstates should have.

While the Gibbons-Hawking hyper-K\"ahler geometries are simple and
``highly computable,'' it is likely that to make significant further
progress in describing and counting the microstates of even ordinary
BMPV black holes (with non-maximal angular momentum), we will have to
get some deeper understanding of the broader class of four-dimensional
generalized hyper-K\"ahler geometries.   An obvious first step is to
look at geometries that still possess a $U(1)$ isometry, but one that
is not tri-holomorphic.  Such geometries are determined by the
$SU(\infty)$ Toda equation \refs{\BoyerMM,\gegenberg}.
In particular, we expect this class of solutions to
describe more general microstates of black holes and black rings, and
perhaps even axially aligned families of such objects.  Such
geometries will also describe the bubbled versions of supertubes with
arbitrary charge densities
\BenaTD\ and perhaps with shape modes that are restricted to a plane.
As we remarked in the introduction, such solutions are also needed
to describe the merging of microstates of two co-axial BMPV black holes.

Even more generally, supersymmetry merely requires that the base be
hyper-K\"ahler, and so the most general, and presumably by far the
most numerous solutions will have no isometries at all.  It would be
useful to see if twistor methods or generalized Legendre
transformations \refs{ \HitchinEA ,\KarlhedeMG,\LindstromKS,
\IvanovCY} could be adapted to this problem.  These methods linearize
the Monge-Amp\'ere equation that underlies the general hyper-K\"ahler
metric, but in practice it is hard to obtain explicit metrics from
this method.  On the other hand, if one is trying to quantize and
count such metrics then this may well be
the natural way to proceed.  Another way to obtain more general families of microstate geometries would be to look for solutions that have a nontrivial metric on the $T^6$, and cannot be reduced to five dimensions \refs{\LuninIZ,\GiustoZI}.
There is evidently a considerable number
of interesting geometries yet to be discovered and studied, and these
geometries lie at the heart of understanding the interior structure of
black holes.

\bigskip
\leftline{\bf Acknowledgments}

We would like to thank Juan Maldacena for interesting discussions. IB
would also like to thank the Santa Barbara KITP and the Aspen Center for Physics for hospitality during various stages at this project. The work of NW and CWW is supported
in part by the DOE grant DE-FG03-84ER-40168.  The work of IB is
supported in part by the NSF grants PHY-0503584 and PHY-990794.

\appendix{A}{More Details on the Entropy of the Foam}

When a foam has a very large number of centers, one can obtain many solutions
with the same charges and angular momenta, but which have different flux
parameters. In \BenaIS\ we estimated the entropy of such arrangements
(which we referred to as ``topological'' entropy) and found that it is proportional to
$Q^{1/4}$. Since all the details of this estimate were not given in \BenaIS\  we
devote this Appendix to this task.

If we work in M theory on $T^6$, then \MtheoryQ\ implies that a black hole foam has
\eqn\sumk{\sum k^1_i = {1 \over 2}  \sqrt{Q_2 Q_3 \over Q_1} ~,}
and similarly for $ \sum k^2_i$ and $\sum k^3_i $. Hence, the total charges do not
depend on the number, GH charges or individual $k_i^I$ of the GH points, but
{\it only} on the sums of the $k_i$.   In \BenaIS\ we found that the non-trivial entropy
coming simply from the many possibilities of choosing the positive, half-integer,
$k_i^1$ subject to the constraint \sumk\ is
\eqn\entropy{S =2 \pi \sqrt{{1\over 6} {\left(Q_2\, Q_3 \over
Q_1\right)} ^{1/2} }\,. }
Naively there should be similar factors coming from partitioning $k^2_i$ and $k^3_i$,
which would lead to a ``topological'' entropy :
\eqn\stotal{S_{topological} = 2  \pi \left( \sqrt{{1\over 6}
{\left(Q_2\, Q_3 \over Q_1\right)} ^{1/2} } +
 \sqrt{{1\over 6} {\left(Q_1 Q_2 \over Q_3\right)} ^{1/2} }
+
 \sqrt{{1\over 6} {\left(Q_1 Q_3 \over Q_2\right)} ^{1/2} }~
\right)~.}

There are, however some subtleties. First, the partitioning of
$k^1_i$, $k^2_i$ and $k^3_i$ is not completely independent.  A bubble
will collapse unless all three fluxes are non-zero, and so we should
count the ways of having non-zero partitions of all the $k^I_i$ over
$N$ bubbles and then sum over $N$.    It is, however, relatively easy to show
that this additional constraint only modifies the entropy \stotal\ by terms that are logarithmic
in the charges.

To see this, it is convenient to define:
\eqn\countfna{g(z,q) ~\equiv~ \prod_{n=1}^\infty {1 \over (1 -z\, q^n)} ~=~ \sum_{N=1}^\infty
d_N(z) \, q^N \,.}
The coefficient of $z^m$ in $d_N(z)$ is the number of (non-zero) partitions
of $N$ into $m$ {\it positive} integers.  The counting function we want is:
\eqn\countfnb{\eqalign{G(w,z; q_1, q_2,q_3) &  ~\equiv~ g(z,q_1) \, g(w,q_2)\,
g(z^{-1} w^{-1},q_3) \cr & ~=~
\sum_{N_1,N_2,N_3=1}^\infty d_{N_1 N_2 N_3}(w,z) \
q_1^{N_1}\,  q_2^{N_2} \, q_3^{N_3}   \,.}}
To count the entropy one must find the coefficient of $w^0 z^0$ in $d_{N_1 N_2 N_3}(w,z)$,
where $N_J = (Q_1 Q_2 Q_3)^{1/2}/Q_J$. Selecting these powers of $w$ and $z$
guarantees that all the partitions are taken over the same number of bubbles.
One can  perform the asymptotic analysis of this partition function  using steepest
descent methods\foot{See, for example, \GreenSP,  pp 116--118.}.  One finds
that for large $N_J$, one has
\eqn\dilogs{d_{N_1 N_2 N_3}(w,z)  ~\sim~ \exp\Big(2\,\big(
\sqrt{N_1\, {\rm Li}_2 (z)} ~+~ \sqrt{N_2\, {\rm Li}_2 (w)} ~+~
\sqrt{N_3\, {\rm Li}_2 (z^{-1} w^{-1})}\,\big) \Big) \,,}
provided that $z,w$ lie in appropriate regions around $z=w=1$.
The functions, ${\rm Li}_2$, are standard dilogarithms.   One can extract the coefficient
of $w^0 z^0$ by contour integrals, but it is easy to see that the leading exponential
behavior of any such integral is dominated by the value of \dilogs\ at $z=w=1$,
and using ${\rm Li}_2(1) = {\pi^2 \over 6}$, one arrives at \stotal.

There is a simple, intuitive way of arriving at this result.  The coefficients, $d_N(z)$,
in \countfna\ are   polynomials of degree $N$  in $z$.
The value  $d_N(1)$ grows as $\exp(2 \pi \sqrt{N/6})$.   Thus, a ``typical coefficient''
in the  polynomial, $d_N(z)$,  grows as  $N^{-1} \exp(2 \pi \sqrt{N/6})$.
Therefore,  restricting to a single power of $z$ simply
leads to $\log(N)$ corrections to the entropy.

The second subtlety is that, given the $k^1_j$,
there are also further constraints on $k^2_j$ and $k^3_j$ imposed by
the global absence of CTC's.  These conditions are somewhat more
difficult to handle but we believe that the bubble equations, combined
with some suitable positivity conditions on the fluxes, will suffice
to guarantee  the conditions in \noCTCs, and hence that \stotal\ is
correct to  leading order.
Independent of these subtleties and the constraints on $k^2_j$ and
$k^3_j$ for a given set of $k^1_j$, we see from \entropy\ alone that
the topological entropy grows as $Q^{1/4}$.

As we discussed in \BenaIS, these calculations indicate that the topological entropy is not enough to account for the entropy of the black hole; to capture the latter one would have to consider microstates that do not reduce to four dimensional multi-center solutions, and that are determined by arbitrary functions \refs{\LuninJY,\LuninIZ,\Saxena}.

\listrefs
\vfill
\eject
\end